%% file: main.tex
\documentclass[sigconf]{acmart}
\AtBeginDocument{%
  \providecommand\BibTeX{{
    \normalfont B\kern-0.5em{\scshape i\kern-0.25em b}\kern-0.8em\TeX}}}

\setcopyright{none}

\acmSubmissionID{75}

\usepackage{pgfplots}
\pgfplotsset{compat=1.12}

\usepackage{nicefrac}
\usepackage{graphicx}
\usepackage{subcaption}
\usepackage{mwe}
\usepackage{multirow}
\usepackage{graphicx}
\usepackage{amsmath}
\usepackage{algorithm}
\usepackage{nicefrac}
\usepackage[export]{adjustbox}
\usepackage{todonotes}

\DeclareMathOperator*{\argmin}{argmin}
\DeclareMathOperator*{\argmax}{argmax}

\newcommand{\dquotes}[1]{``#1''}

\newif\iffair
\fairfalse

\definecolor{blue(pigment)}{rgb}{0.2, 0.2, 0.7}

\def\ml{\texttt{ML-1M}\xspace}
\def\lfm{\texttt{LastFM}\xspace}

\begin{document}

\title{Multi-Step Adversarial Perturbations on Recommender Systems Embeddings}

\author{Vito Walter Anelli}
\email{vitowalter.anelli@poliba.it}
\affiliation{%
  \institution{Polytechnic of Bari, Italy}
}

\author{Alejandro Bellogín}
\email{alejandro.bellogin@uam.es}
\affiliation{%
  \institution{University of Madrid, Spain}
}

\author{Yashar Deldjoo}
\orcid{0000-0002-6767-358X}
\email{yashar.deldjoo@poliba.it}
\affiliation{%
  \institution{Polytechnic of Bari, Italy}
}

\author{Tommaso Di Noia}
\email{tommaso.dinoia@poliba.it}
\affiliation{%
  \institution{Polytechnic of Bari, Italy}
}

\author{Felice Antonio Merra}
\authornote{Authors are listed in alphabetical order. Corresponding author: Felice Antonio Merra.}
\email{felice.merra@poliba.it}
\affiliation{%
  \institution{Polytechnic of Bari, Italy}
}

\begin{abstract}
\input{0Abstract}
\end{abstract}

\keywords{Adversarial Machine Learning, Recommender Systems}

\maketitle

\fancyhead[]{}
\input{1Introduction}

\section{Related Work}\label{sec:related}
\input{5Background}

\section{The Proposed Framework}\label{sec:proposed_framewok}
\input{2ProblemFormalization}

\section{Experimental Setup}\label{sec:experiment}
\input{3Experiment}

\input{images/figures}
\section{Results and Discussion}\label{sec:discussion}
\input{4Discussion}

\section{Conclusion and Future Work}\label{sec:conclusion}
\input{6Conlusion}


\bibliographystyle{ACM-Reference-Format}
\bibliography{main}

\end{document}

%% file: 0Abstract.tex
Recommender systems (RSs) have attained exceptional performance in learning users’ preferences and helping them in finding the most suitable products. Recent advances in adversarial machine learning (AML) in the computer vision domain have raised interests in the security of state-of-the-art model-based recommenders. Recently, worrying deterioration of recommendation accuracy has been acknowledged on several state-of-the-art model-based recommenders (e.g., BPR-MF) when machine-learned adversarial perturbations contaminate model parameters. However, while the single-step fast gradient sign method (FGSM) is the most explored perturbation strategy, multi-step (iterative) perturbation strategies, that demonstrated higher efficacy in the computer vision domain, have been highly under-researched in recommendation tasks.

In this work, inspired by the basic iterative method (BIM) and the projected gradient descent (PGD) strategies proposed in the CV domain, we adapt the multi-step strategies for the item recommendation task to study the possible weaknesses of embedding-based recommender models under minimal adversarial perturbations. 
Letting the magnitude of the perturbation be fixed, we illustrate the highest efficacy of the multi-step perturbation compared to the single-step one with extensive empirical evaluation on two widely adopted recommender datasets. Furthermore, we study the impact of structural dataset characteristics, i.e., sparsity, density, and size, on the performance degradation issued by presented perturbations to support RS designer in interpreting recommendation performance variation due to minimal variations of model parameters. Our implementation and datasets are available at \url{https://anonymous.4open.science/r/9f27f909-93d5-4016-b01c-8976b8c14bc5/}.

%% file: 1Introduction.tex
\section{Introduction}\label{sec:introduction}

Machine learning (ML) models play an essential role in the everyday lives of people. Recommender systems (RSs) are involved in many decision-making and mission-critical tasks; thus, the quality of learned models could directly impact people's lives, society, and businesses. The fundamental assumption underneath trained recommender models is that user-item historical interactions can serve as the appropriate ground-truth to learn users' preferences. Consequently, machine-learned recommendation models, such as latent-factor models (LFMs), try to learn what users might like by using any available signal in the data that can minimize an empirical loss, point-wise, pair-wise, or even list-wise~\cite{DBLP:books/sp/Aggarwal16,DBLP:conf/uai/RendleFGS09,DBLP:journals/corr/HidasiKBT15}. However, users' feedbacks are imperfect due to either biases~\cite{DBLP:journals/ir/BelloginCC17} or misoperations. An adversarial perturbation mechanism may learn and leverage imperfections and approximations made by the ML model during the training phase to guide the recommendation outcomes toward an engineered --- and often illegitimate --- purpose, e.g., to reduce the trust of an e-commerce competitor platform.

Recent advances in the study of the security of ML models, named \textit{adversarial machine learning} (AML)~\cite{DBLP:conf/ccs/HuangJNRT11}, have revealed the vulnerability of state-of-the-art ML models for tasks across variety of domains, such as computer vision (CV)~\cite{DBLP:journals/access/AkhtarM18}, natural language processing (NLP)~\cite{10.1145/3374217}, and more recently RS~\cite{DBLP:journals/corr/abs-2005-10322}. Adversarial attacks against ML models were formalized for the first time in the pioneering work by Szegedy et al.~\cite{DBLP:journals/corr/SzegedyZSBEGF13} in which the authors  evaluated the vulnerability of state-of-the-art deep neural networks for \textit{image classification} against adversarial perturbation of the input image data (pixel values). They found that a modest perturbation level is sufficient to confuse the system to classify a given instance of pandas image into the wrong gibbon class with high confidence, whereby human judgment is unaffected. Adversarial perturbations are generated by a model, named \textit{the adversarial attack} model, whose goal is to add a norm-constrained amount of perturbation on the data ($x = x + \Delta$ where $\left\lVert \Delta \right\rVert < \epsilon$), to enforce the ML model to make a wrong prediction (e.g., misclassification). \textit{Adversarial defense} is the design of mechanisms that can withstand and mitigate the impact of adversarial attacks --- i.e., they aim to enhance the \textit{robustness} against adversarial threats.

In the field of CV, several adversarial attack strategies have been proposed such as FGSM~\cite{DBLP:journals/corr/GoodfellowSS14}, BIM~\cite{DBLP:conf/iclr/KurakinGB17a}, PGD~\cite{DBLP:conf/iclr/MadryMSTV18}, and Carlini and Wagner~\cite{DBLP:conf/sp/Carlini017} as well as defense mechanisms such as adversarial training~\cite{DBLP:journals/corr/GoodfellowSS14} and knowledge distillation~\cite{DBLP:journals/corr/HintonVD15, DBLP:conf/sp/PapernotM0JS16}. Novel powerful attack methods are followed by the defensive side of this war, in which stronger defense strategies are proposed to confront these attacks and to reduce their impact.

However, attack and defense strategies are treated differently in the CV and RS communities due to intrinsic differences between images and RS data. For instance, image data (pixel intensities) is continuous-valued but RSs data is discrete in nature such as rating scores or binary-valued interactions (either 0 or 1). For this reason, the mainline RS research injects additive adversarial perturbations on model parameters (e.g.,  embedding matrix of an LFM), instead of raw pixel intensities as in for images, and evaluates the variation of recommendation performance caused by a small and optimized adversarial noise

The pioneering work of AML for the item recommendation task has been presented by He et al. in~\cite{DBLP:conf/sigir/0001HDC18}. The authors compared the impact of \textbf{adversarial perturbation} generated randomly and via an adversarial model against Bayesian Pairwise Ranking matrix factorization (BPR-MF)~\cite{DBLP:conf/uai/RendleFGS09}, the state-of-the-art pairwise recommendation model. They discovered that the adversarial noise obtained from a \textit{single-step adversarial perturbation} model~\cite{DBLP:journals/corr/GoodfellowSS14} leads to a significant deterioration of test accuracy, which is 5 times larger than the one caused by random variation, i.e., -10\% v.s. -55\% reduction of $nDCG@100$ for a given fixed perturbation budget ($\epsilon=1$). These results raise the question: \dquotes{\textit{how much state-of-the-art machine-learned models are robust against small adversarial perturbations?}}. Note that the application scenario of these adversarial perturbations is not related to an adversary that enters in an RS and changes the parameters of the model. In fact, as proposed by He et al.~\cite{DBLP:conf/sigir/0001HDC18}, 
it aims to study the weaknesses when training a recommender model that might learn parameters such that 
a small variation will cause a drastic performance degradation. For instance, this variation might be potentially caused by the update of model parameters when new users, or products, are added into the system.

Furthermore, He et al.~\cite{DBLP:conf/sigir/0001HDC18} demonstrated that by applying an \textit{adversarial training} procedure~\cite{DBLP:journals/corr/GoodfellowSS14}, named adversarial personalized ranking (APR), the RS performs more robustly against adversarial perturbations generated by the single-step method and might improve the model generalizability with a consequent \textit{improvement in accuracy}~\cite{DBLP:conf/sigir/0001HDC18, DBLP:conf/sigir/0001HDC18, DBLP:conf/sigir/YuanYB19}. However, it is not studied how brittle AMF --- i.e., the APR-robustified version of BPR-MF--- and BPR-MF are against perturbations generated by other adversarial strategies.  In fact, despite the popularity of AML for security in ML/CV and recently in RS communities, we have noted that there exists a \textit{pronounced imbalance} in the distribution of research efforts. Among most papers studied in recent years~\cite{DBLP:journals/corr/abs-2005-10322}, almost all RS papers specialize in one attack model, namely FGSM~\cite{DBLP:journals/corr/GoodfellowSS14, DBLP:conf/sigir/0001HDC18}.
However, in the CV domain, \textit{iterative adversarial perturbations} have been demonstrated to improve the attack effectiveness by more than 60\% compared to FGSM~\cite{DBLP:conf/iclr/KurakinGB17a}. To the best of our knowledge, no major attempt has been made in the RS community to study the RS performance variation when model embeddings are altered by multi-step perturbations.

To fill this gap, in this work, we present two CV-inspired multi-step adversarial perturbation mechanisms, namely the basic iterative method (BIM)~\cite{DBLP:conf/iclr/KurakinGB17a} and projected gradient descent (PGD)~\cite{DBLP:conf/iclr/MadryMSTV18}. We empirically investigate the robustness of the existing defense strategy in RSs, APR applied on BPR-MF, against the proposed perturbations. We focus our study on adversarial perturbation on BPR-MF based models motivated by the fact that the adversarial perturbation and defense mechanisms proposed by He et al.~\cite{DBLP:conf/sigir/0001HDC18} have been designed for BPR-MF and then applied in several recommender models (e.g., AMR~\cite{8618394}, FGACAE~\cite{DBLP:conf/sigir/YuanYB19}, SACRA~\cite{DBLP:conf/wsdm/LiW020}).

To provide better understanding of the proposal, we run extensive experiments to study the impact of data characteristics such as data sparsity on attack/defense methods' performance and evaluate the performance of the system against beyond-accuracy metrics, i.e., expected free discovery, Shannon Entropy, and item coverage.
This work aims to answer the following research questions:
\vspace{-0.9mm}
\begin{itemize}
    \item Against the common background in the field of RS, that recognizes FGSM as the main adversarial perturbation strategy, and \textit{let the perturbation budget be fixed}, do the multi-step adversarial perturbation models, PGD and BIM, outperform FGSM in degrading the quality of system, with respect to accuracy and beyond-accuracy evaluation measures? 
    
    \item
    Is the state-of-the-art defensive mechanism, adversarial personalized ranking (APR), still useful against the presented iteratively generated noise?
    
    \item
    Structural user-item data characteristics have been demonstrated to influence the effectiveness of shilling attacks~\cite{sisinflabSigir2020}. Is this influence confirmed also in the case of $\epsilon$-bounded adversarial perturbation of model parameters?

\end{itemize}

To this end, we evaluated the proposed strategies against two standard model-based collaborative recommenders, i.e., BPR-MF~\cite{DBLP:conf/uai/RendleFGS09} and its adversarial robustified version AMF~\cite{DBLP:conf/sigir/0001HDC18}, on two well-recognized recommender datasets, i.e., \ml and \lfm. Overall, the considered strategies highlight the necessity to investigate new robustification methods to make the recommender models more robust to the performance worsening caused by this minimal noise.



%% file: 5Background.tex
In this section, we present background on collaborative recommenders and the advances of adversarial machine learning in recommendation settings.


\subsection{Collaborative Recommendation.}
Different types of recommendation models have been proposed in the last thirty years, which can be mainly categorized into collaborative filtering (CF), content-based, and hybrid RS~\cite{DBLP:reference/sp/2015rsh}. The first category learns users' preferences from historical user-item interactions, e.g., User- and  Item-\textit{k}NN ~\cite{DBLP:journals/tkdd/Koren10} and BPR-FM~\cite{DBLP:conf/uai/RendleFGS09}. The second category suggests unseen products based on the content-based similarity of user consumed items and other unseen items~\cite{DBLP:reference/sp/GemmisLMNS15}. The last class combines both techniques to augment user-item interactions with side information~\cite{DBLP:journals/umuai/Burke02}. Model-based CF models such as BPR-MF~\cite{DBLP:conf/uai/RendleFGS09} and recent neural models such as neural collaborative filtering (NCF)~\cite{DBLP:conf/www/HeLZNHC17} are popular choices in the RS and ML communities. BPR-MF is the major recommendation model used in the RS community for research on adversarial attacks~\cite{DBLP:journals/corr/abs-2005-10322}. In this work, we chose BPR-MF and AMF~\cite{DBLP:conf/sigir/0001HDC18}, that applies an adversarial training on BPR-MF (see Eq.~\ref{eq:l_amf}), as the core recommendation methods tested for all the conducted experiments. 




\subsection{Adversarial Machine Learning in Recommender Systems.} Adversarial Machine Learning (AML) approaches in RSs are classified based on their application on either (i) the security of RS or, (ii) the learning model of the generative adversarial networks (GANs)~\cite{DBLP:journals/corr/GoodfellowPMXWOCB14} in GAN-based RS~\cite{DBLP:journals/corr/abs-2005-10322}. Security (i.e., attack on and defense of)
RS has been studied in two lines of research~\cite{DBLP:journals/corr/abs-2005-10322}: based on hand-engineered shilling attack and, (ii) machine-learned perturbations using AML techniques. The former category is based on the injection of fake profiles manually generated by malicious users~\cite{DBLP:reference/sp/BurkeOH15}. The latter category, which is the focus of the current work, studied the application of ML techniques to generate optimal perturbations to reduce the performance of a recommender model~\cite{DBLP:conf/sigir/0001HDC18, DBLP:conf/recsys/Christakopoulou19, DBLP:conf/wsdm/BeigiMGAN020} and their defense countermeasures~\cite{DBLP:conf/sigir/0001HDC18, DBLP:conf/wsdm/EntezariADP20}. Pioneering work by He \textit{et al.}~\cite{DBLP:conf/sigir/0001HDC18} reported grave vulnerability of BPR-MF against adversarial perturbation obtained from the fast gradient sign method (FGSM) attack model and suggested an adversarial training procedure, named \textit{adversarial regularization} as a defensive countermeasure. Inspired by this work, several other works incorporated adversarial training procedure in several recommendation tasks and models e.g., AMR~\cite{8618394} on fashion recommendation, FG-ACAE~\cite{DBLP:conf/sigir/YuanYB19, DBLP:conf/ijcnn/YuanYB19} for collaborative deep recommendation, ATF~\cite{DBLP:conf/recsys/ChenL19} for tensor factorization and so forth. However, we found that research in RS community lacks sufficient relevant studies on another category of adversarial attacks such as iterative attacks, e.g., BIM~\cite{DBLP:conf/iclr/KurakinGB17a} and PGD~\cite{DBLP:conf/iclr/MadryMSTV18}, which have been shown to be effective in tampering computer vision tasks. This work provides an exhaustive analysis of this gap between the lack of works using multi-step adversarial perturbation in the RS domain in order to investigate if the recommendation performance are much worse with iterative perturbations than with non-iterative ones by fixing the size of the adversarial noise.

%% file: 2ProblemFormalization.tex
In this section, we describe the foundations of a personalized matrix factorization (MF) recommender model. Then, we recapitulate the baseline single-step adversarial perturbation before defining the multi-step strategies.

\subsection{Personalized Recommenders via Matrix Factorization.}
The recommendation problem is the task of estimating a preference prediction function $s(u, i)$ that maximizes the utility of the user $u \in \mathcal{U}$ in getting the item $i \in \mathcal{I}$ recommended by the RS, where $ \mathcal{U}$ and  $\mathcal{I}$ are the set of users and items respectively.
Before we dive into the description of the matrix factorization model, we introduce the following notation:
\begin{itemize}
    \item $\mathbf{P}$: the matrix of \textit{user} embeddings, where $\mathbf{p}_u$ is the embedding vector associated to the user $u$;
    \item $\mathbf{Q}$: the matrix of \textit{item} embeddings, where $\mathbf{q}_i$ is the embedding vector associated to the item $i$;
    \item $\mathbf{\Theta}$: the set of model parameters ($\mathbf{\Theta} = \{\mathbf{P}, \mathbf{Q}\}$);
    \item $\mathcal{L}$: the loss function
\end{itemize}

The main intuition behind the MF model is to compute the preference score $s(u,i)$ as the dot product between the user's embedding ($\mathbf{p}_u$) and the item's embedding ($\mathbf{q}_i$), i.e., $s(u, i) = \mathbf{p}_u^T \mathbf{p}_i$. Bayesian Personalized Ranking (BPR)~\cite{DBLP:conf/uai/RendleFGS09} that uses a pairwise learning-to-rank method, is the state-of-the-art approach to produce personalized rankings. BPR is based on the idea of reducing ranking problem to pairwise classification problem between interacted and non-interacted products. The model parameters are learned by solving the optimization problem in the following general form:
\begin{equation}
    \argmin_{\mathbf{\Theta}} \mathcal{L(\mathbf{\Theta})}
    \label{eq:minim}
\end{equation}
To solve the minimization problem in Eq.~\ref{eq:minim}, in our experimental evaluation, we used the adaptive gradient optimizer (Adagrad).




\subsection{Adversarial Perturbation of Model Parameters.}
The main intuition behind an adversarial perturbation method is to generate minimum perturbations ($\mathbf{\Delta}^{adv}$) capable of undermining the learning objective of the learning model. The adversary’s goal is to maximize Eq.~\ref{eq:minim}, under a minimal-norm constraint:
\begin{equation}
    \Delta^{adv} \leftarrow \argmax_{\footnotesize{\mathbf{\Delta}_0,||\mathbf{\Delta}_{0}|| \leq \epsilon}} \mathcal{L}(\mathbf{\Theta}+\mathbf{\Delta}_0)\label{eq:maxim}
\end{equation}
where $\mathbf{\Delta}_0$ is the initial adversarial perturbation added to the model parameters $\mathbf{\Theta}$ at the beginning of the adversarial perturbation and  $\epsilon$ is the \textit{perturbation budget}, the limit the amount of perturbation. Equations~\ref{eq:minim} and~\ref{eq:maxim} can be represented under unique \textit{minimax} optimization formulation presented below:
\begin{equation}
    \arg \quad \min_{\mathbf{\Theta}} \max _{\footnotesize{\mathbf{\Delta}_0,||\mathbf{\Delta}_{0}|| \leq \epsilon}} \mathcal{L}(\mathbf{\Theta}+\mathbf{\Delta}_0)
    \label{eq:minimax}
\end{equation}
in which two opposite players play an adversarial minimax game, where the adversary tries to maximize the likelihood of its success while the ML model tries to minimize the risk. This minimax game is the main characteristic of tasks related to AML research~\cite{DBLP:conf/nips/2019, DBLP:journals/corr/abs-2005-10322}.

\subsubsection{Single-Step Adversarial Perturbation (FGSM)}~\cite{DBLP:conf/sigir/0001HDC18}. This perturbation strategy is the baseline single-step adversarial noise mechanism to alter item recommendation results and was first proposed by He et al. in~\cite{DBLP:conf/sigir/0001HDC18}. It builds on advanced made in ML research  pioneered by Goodfellow et al.~\cite{DBLP:journals/corr/GoodfellowSS14} for the classification task. It approximates $\mathcal{L}$ by linearizing it around an initial zero-matrix perturbation $\mathbf{\Delta}_0$ and applies the max-norm constraint. The adversarial perturbation $\mathbf{\Delta}^{adv}$ is defined as:
\begin{equation}
    \mathbf{\Delta}^{adv}=\epsilon \frac{\Pi}{\|\Pi\|} \quad \text { where } \quad \Pi=\frac{\partial \mathcal{L}(\mathbf{\Theta}+ \mathbf{\Delta}_0)}{\partial \mathbf{\Delta}_0}
    \label{eq:fgdm-he}
\end{equation}
where $||\cdot||$ is the $L_2-$norm. After the calculation of $\mathbf{\Delta}^{adv}$, Goodfellow et al. added this perturbation to the current model parameters $ \mathbf{\Theta}^{adv} = \mathbf{\Theta} + \mathbf{\Delta}^{adv}$ and generated the recommendation lists with this perturbed model parameter. He et al. in~\cite{DBLP:conf/sigir/0001HDC18} demonstrated that perturbation obtained from the FGSM with $\epsilon=0.5$ can impair the accuracy of item recommendation by an amount equal to $-26.3\%$.

\subsubsection{Multi-Step Adversarial Perturbation.} This adversarial noise generation mechanism is a straightforward extension of the single-step strategy and was proposed by Kurakin et al.~\cite{DBLP:conf/iclr/KurakinGB17a}. In particular, the authors' idea was to build an FGSM-based \textit{multi-step} strategy and creating more effective $\epsilon$-clipped perturbations. 
The initial model parameters are defined as 
\begin{equation}
    \mathbf{\Theta} ^{adv}_{0} = \mathbf{\Theta} + \mathbf{\Delta}_0
    \label{eq:initialization}
\end{equation}
Starting from this initial state of model parameters, let $Clip_{\mathbf{\Theta} , \epsilon}$ be an element-wise clipping function to limit the perturbation of each original embedding value inside the $[-\epsilon, +\epsilon]$ interval, let $\alpha$ be the step size which is the maximum perturbation budget of each iteration, and let $L$ be the number of iterations, the first iteration ($l=1$) is defined by: 
\begin{equation}
    \mathbf{\Theta} ^{adv}_{1} = Clip_{\mathbf{\Theta} , \epsilon} \left\{ \boldsymbol{\mathbf{\Theta} }_{0}^{a d v}+\alpha  \frac{\Pi}{\| \Pi \|} \right\} \text{ where } \Pi = \frac{\partial \mathcal{L}(\mathbf{\Theta} + \mathbf{\Delta}_0)}{\partial \mathbf{\Delta}_0 }
\end{equation}
and we generalize the $l$-th iteration of the $L$-iterations multi-step adversarial perturbation as: 
\begin{equation}
    \mathbf{\Theta} ^{adv}_{l} = Clip_{\mathbf{\Theta} , \epsilon}\left\{\boldsymbol{\mathbf{\Theta} }_{l-1}^{a d v}+\alpha  \frac{\Pi}{\| \Pi \|} \right\} \text{ where } \Pi = \frac{\partial \mathcal{L}(\mathbf{\Theta} + \mathbf{\Delta}^{adv}_{l-1})}{\partial \mathbf{\Delta}^{adv}_{l-1} }
\end{equation}
where $l \in [1,2, ..., L]$, $\mathbf{\Delta}^{adv}_l$ is the adversarial perturbation at the $l$-th iteration, and $\mathbf{\Theta} ^{adv}_{l}$ is the sum of the original model parameters $\mathbf{\Theta}$ with the perturbation at the $l$-th iteration. Inspired by the advances of AML in iterative adversarial perturbations, we considered two different versions of multi-step optimized adversarial perturbation: the Basic Iterative Method (\textbf{BIM})~\cite{DBLP:conf/iclr/KurakinGB17a} and the Projected Gradient Descent (\textbf{PGD})~\cite{DBLP:conf/iclr/MadryMSTV18} approaches. 
The former approach initializes $\mathbf{\Delta}_0$ of Eq.~\ref{eq:initialization} as zeros matrices with the same size of the matrix embeddings of the victim model (the model under perturbation), i.e., $\mathbf{P}$ and $\mathbf{Q}$. The latter is the BIM extension that sets the initial perturbation by sampling it from a uniform distribution. This difference in the initialization of $\mathbf{\Delta}_0$ makes PGD more powerful than BIM in confusing CV image classifiers~\cite{DBLP:conf/icml/AthalyeC018}. We chose both strategies to investigate whether such a difference between two adversarial perturbation strategies exists for the recommendation task.



%% file: 3Experiment.tex
In this section, we introduce the two explored datasets, the model-based RSs, and the set of evaluation measures. Then, we detail the experimental choices to make the results reproducible.

\subsection{Datasets}
To experiment with our proposed iterative perturbations, we conducted experiments on two public datasets: 

\textbf{Movielens 1M} (\ml)~\cite{DBLP:journals/tiis/HarperK16} contains users' explicit feedbacks (i.e., ratings) towards a catalog of movies given in the [1, 5]-scale. We took the public version within the set of users' additional attributes (e.g., user's gender, age, and occupation) and movies' genres.  The dataset has 6,040 users ($|\mathcal{U}|$), 3,706 items ($|\mathcal{I}|$), and 1,000,209 recorded feedbacks ($|\mathcal{F}|$).

\textbf{LastFM-1b} (\lfm)~\cite{DBLP:conf/mir/Schedl16} contains more than one billion recorded interactions (e.g., the user's actions of listening tracks) stored from the online music provider Last.fm from January 2013 to August 2014. We used the pre-processed version proposed in~\cite{sisinflabSigir2020}, extracting a sampled version with several recorded feedbacks comparable to the \ml ones (i.e., about 1 million). The experimented dataset has 2,847 users ($|\mathcal{U}|$), 33,164 items ($|\mathcal{I}|$), and 935,875 historical preferences ($|\mathcal{F}|$).

For both the datasets, we transformed all the recorded interactions into 1-valued implicit feedback following the dataset experimental setting proposed in~\cite{DBLP:conf/sigir/0001HDC18}. 

\subsection{Recommender Models}\label{subsubsec:rec}

We verified the impact of the proposed iterative adversarial perturbations against two baseline recommender models: BPR-MF~\cite{DBLP:conf/uai/RendleFGS09} and AMF~\cite{DBLP:conf/sigir/0001HDC18}. In particular, the second model has been specifically proposed to make BPR-MF more robust under the single-step adversarial perturbation (FGSM).

\textbf{BPR-MF}~\cite{DBLP:conf/uai/RendleFGS09} is a matrix factorization recommender optimized with a pair-wise loss function (i.e., BPR). The fundamental intuition of BPR-MF is to discard not-interacted items with respect to interacted ones in order to learn a rank-based preference predictor. We denote with $\mathcal{L}_{BPR}(\mathbf{\Theta}) = \mathcal{L}(\mathbf{\Theta})$ the BPR-MF loss function.

\textbf{AMF}~\cite{DBLP:conf/sigir/0001HDC18} is the extension of BPR-MF which includes an adversarial training procedure. The main idea is to extend the BPR-MF loss function within a classical adversarial \textit{mini-max} game. In fact, the authors proposed to include additional training steps to minimize the following loss function:
\begin{equation}
    \label{eq:l_amf}
    \mathcal{L}_{AMF}(\mathbf{\Theta}) = \mathcal{L}_{BPR}(\mathbf{\Theta}) + \lambda \underbrace{\mathcal{L}_{BPR}(\mathbf{\Theta}^{adv})}_{\text{adversarial regularizer}}
\end{equation}
where the \textit{adversarial regularizer} component builds the adversarial perturbation ($\mathbf{\Theta}^{adv}$) with the FGSM approach described in Eq.~\ref{eq:fgdm-he}. This model has been demonstrated to reduce up to 88\% the impact of single-step perturbations on the recommendation accuracy~\cite{DBLP:conf/sigir/0001HDC18}.

\subsection{Evaluation Metrics}
We conducted a three-level analysis to answer the research questions defined in Section~\ref{sec:introduction}. 
In the first level we focused on assessing the efficacy of our proposed iterative perturbations under the lens of the destruction of the recommendation accuracy. To handle this objective we studied precision ($PR@K$), recall ($RE@K$), and normalized discounted cumulative gain ($nDCG@K$) evaluated on per-users' top-$K$ recommendation lists~\cite{DBLP:reference/sp/GunawardanaS15}. 
$PR@K$ measures the fraction of suggested items relevant to the users concerning the length of the recommendations ($K$), $RE@K$ evaluates the average fraction of user's relevant recommended items in the top-$K$ over the number of items in her test set, and $nDCG@K$ computes the users' utility (\textit{gain}) given the ranked list by discounting the correctly predicted relevant items by their positions. 

\iffalse
Furthermore, we explored the effect of iterative perturbations with respect to beyond accuracy metrics. In particular, we analyzed three metrics: the expected free discovery ($EFD@K$), the Shannon Entropy ($SE@K$), and the item coverage ($ICov@K$).
The first one ($EFD@K$) measures the \textit{novelty} of the RS defined as the capacity to suggest relevant long-tail products~\cite{DBLP:conf/recsys/VargasC11}. The second one ($SE@K$) measures the diversity in the recommendation lists, it is close to 0 when few items are always recommended, while it increases when all items are equally recommended~\cite{DBLP:reference/sp/TintarevM15}. The last metric calculates the number of recommended products across all the top-$K$ users' recommendation lists~\cite{DBLP:conf/recsys/GeDJ10}.
\else
Furthermore, we explored the effect of iterative perturbations with respect to beyond accuracy metrics. In particular, we analyzed three metrics: \textit{novelty} by computing the capacity to suggest relevant long-tail products using the expected free discovery ($EFD@K$)~\cite{DBLP:conf/recsys/VargasC11}, the \textit{diversity} in the recommendation lists using the Shannon Entropy ($SE@K$)~\cite{DBLP:reference/sp/TintarevM15}, and the item \textit{coverage} ($ICov@K$) as the number of recommended products across all the top-$K$ users' recommendation lists~\cite{DBLP:conf/recsys/GeDJ10}.
\fi

\subsection{Reproducibility}
In this section, we provide reproducibility details. 

\textbf{Evaluation Protocol.} The evaluation protocol employed to verify the proposed iterative adversarial perturbations is the leave-one-out protocol~\cite{DBLP:conf/uai/RendleFGS09, DBLP:conf/sigir/0001HDC18}, putting in the test set either the last historical interaction --- when that information is available (i.e., \ml)-- or a random interaction (i.e., \lfm), and using the rest of the recorded feedbacks to train the recommenders. 
Furthermore, we evaluated the generated top-$10$ recommendation lists to get results closer to a real-world exposure of recommendations to customers, filtering out the already interacted products for each user's list.

\textbf{Recommender Models.}
We followed the training scheme proposed in~\cite{DBLP:conf/sigir/0001HDC18}. We trained the defense-free recommender model, BPR-MF, for 2,000 epochs with the model embedding size fixed to $64$, the learning rate values to $0.05$, without regularization coefficients. Furthermore, we stored the model parameters at the 1,000 epoch to be the starting point for the adversarial regularized version of BPR-MF, i.e.,  AMF (see Section~\ref{subsubsec:rec}). We set the adversarial regularization coefficient ($\lambda$) to $1$ as suggested by He et al.~\cite{DBLP:conf/sigir/0001HDC18}. Starting from the restored BPR-MF model, we trained AMF for additional 1,000 epochs. We optimized both the recommenders using the mini-batch adaptive gradient optimizer (Adagrad) with a batch size fixed to 512.  Moreover, we have compared the performance of the adversarially perturbed model-based recommender with the performance of the random recommender.

\textbf{Adversarial perturbations.} For the baseline FGSM perturbation, we fixed the budget perturbation $\epsilon$ to $0.5$, which is the smallest perturbation experimented in~\cite{DBLP:conf/sigir/0001HDC18}. Then, we implemented BIM and PGD setting the step size $\alpha$ to $\nicefrac{\epsilon}{4}$ and varying the budget perturbation ($\epsilon$) and the number of iterations ($L$) depending on the research question under investigation. 

%% file: images/figures.tex
\begin{figure*}[!t]

        \centering
        \begin{adjustbox}{minipage=\linewidth,scale=0.75}
        \begin{subfigure}[b]{0.45\textwidth}
            \centering
            \includegraphics[width=\textwidth]{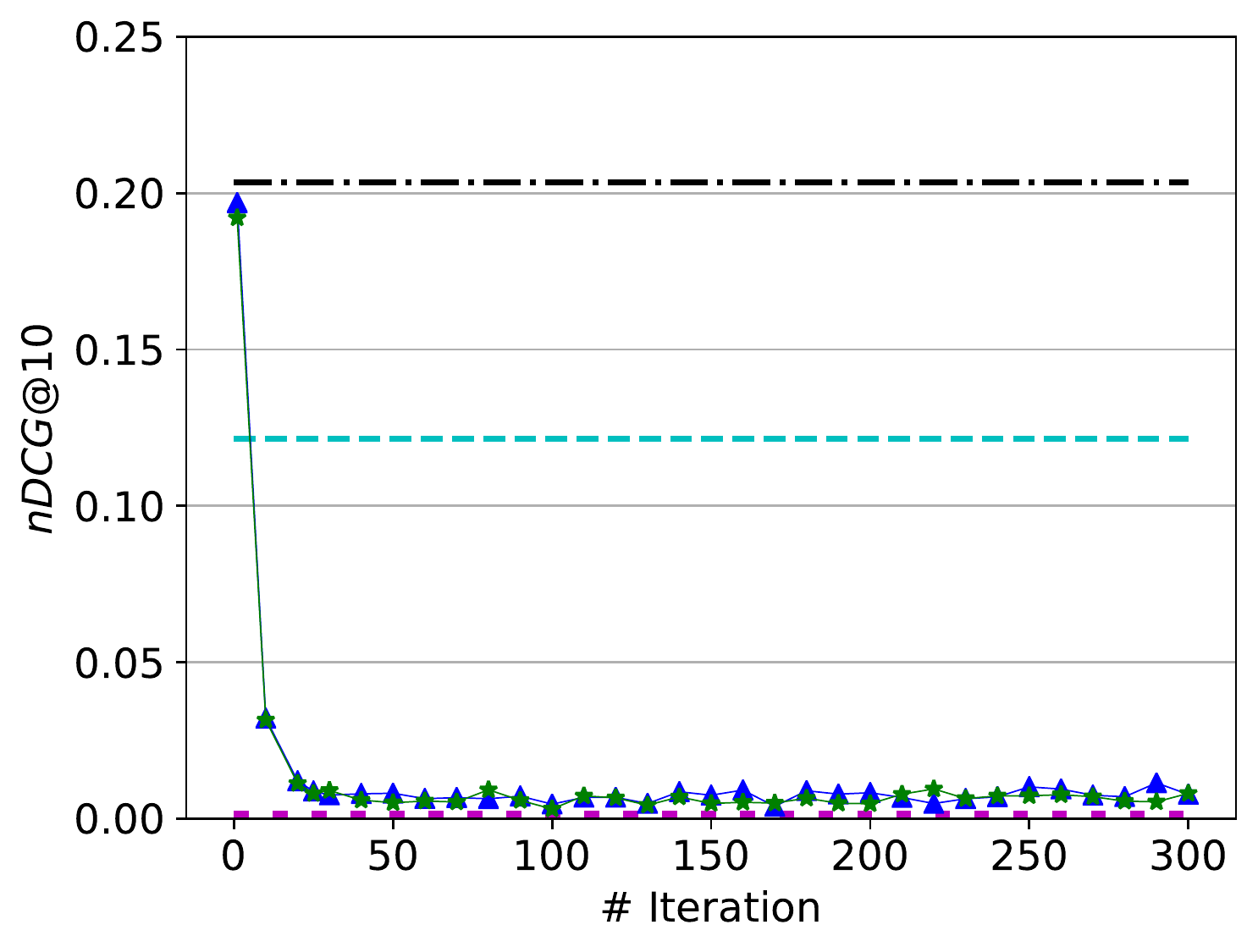}
            \caption{\small Defense-free \textbf{BPR-MF} model.}
            \label{fig:a}
        \end{subfigure}
        \begin{subfigure}[b]{0.45\textwidth}   
            \centering 
            \includegraphics[width=\textwidth]{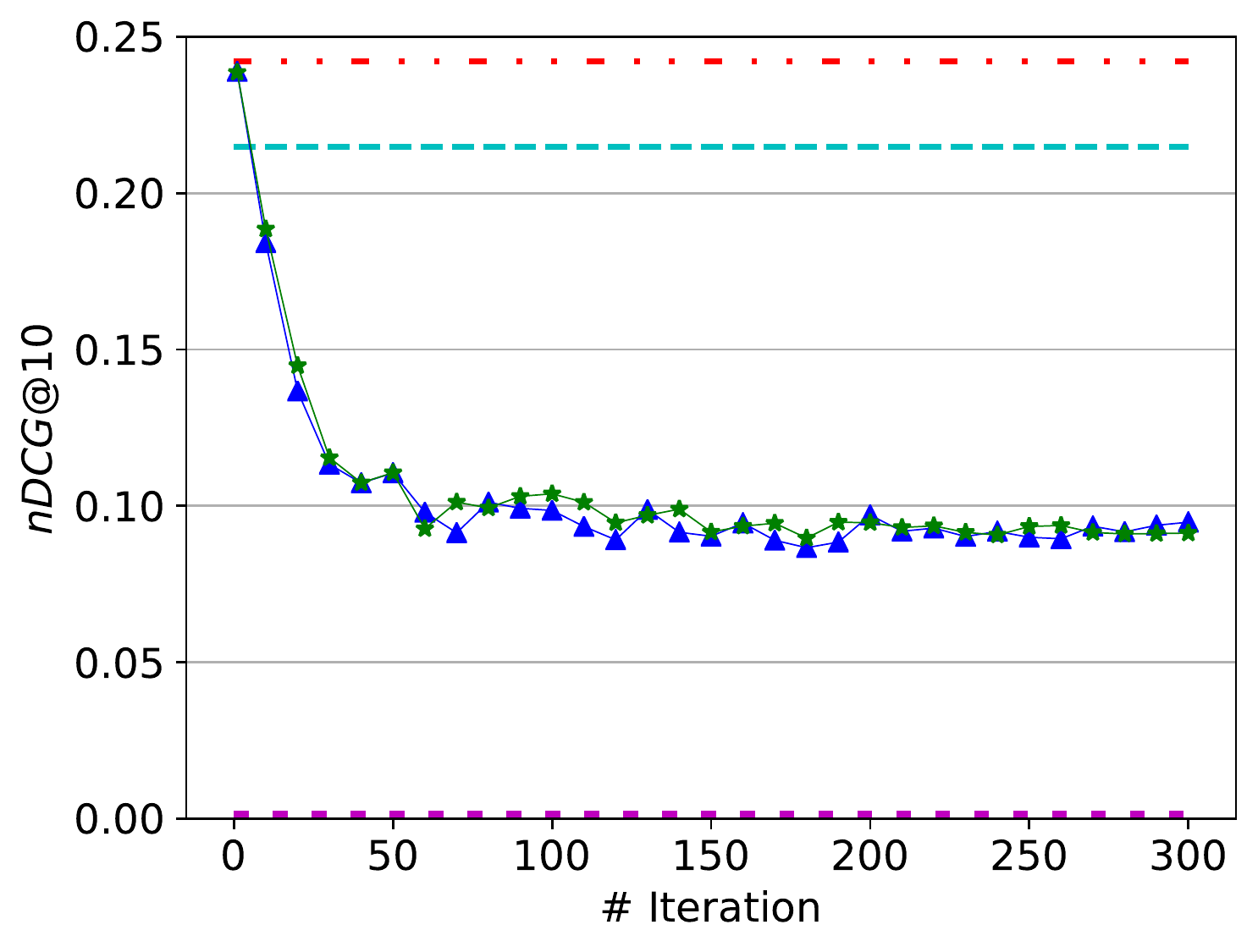}
            \caption{\small Defended BPR-MF model - \textbf{AMF}.}
            \label{fig:b}
        \end{subfigure}
        
        \vskip\baselineskip
        
        \begin{subfigure}[b]{0.45\textwidth}
            \centering
            \includegraphics[width=\textwidth]{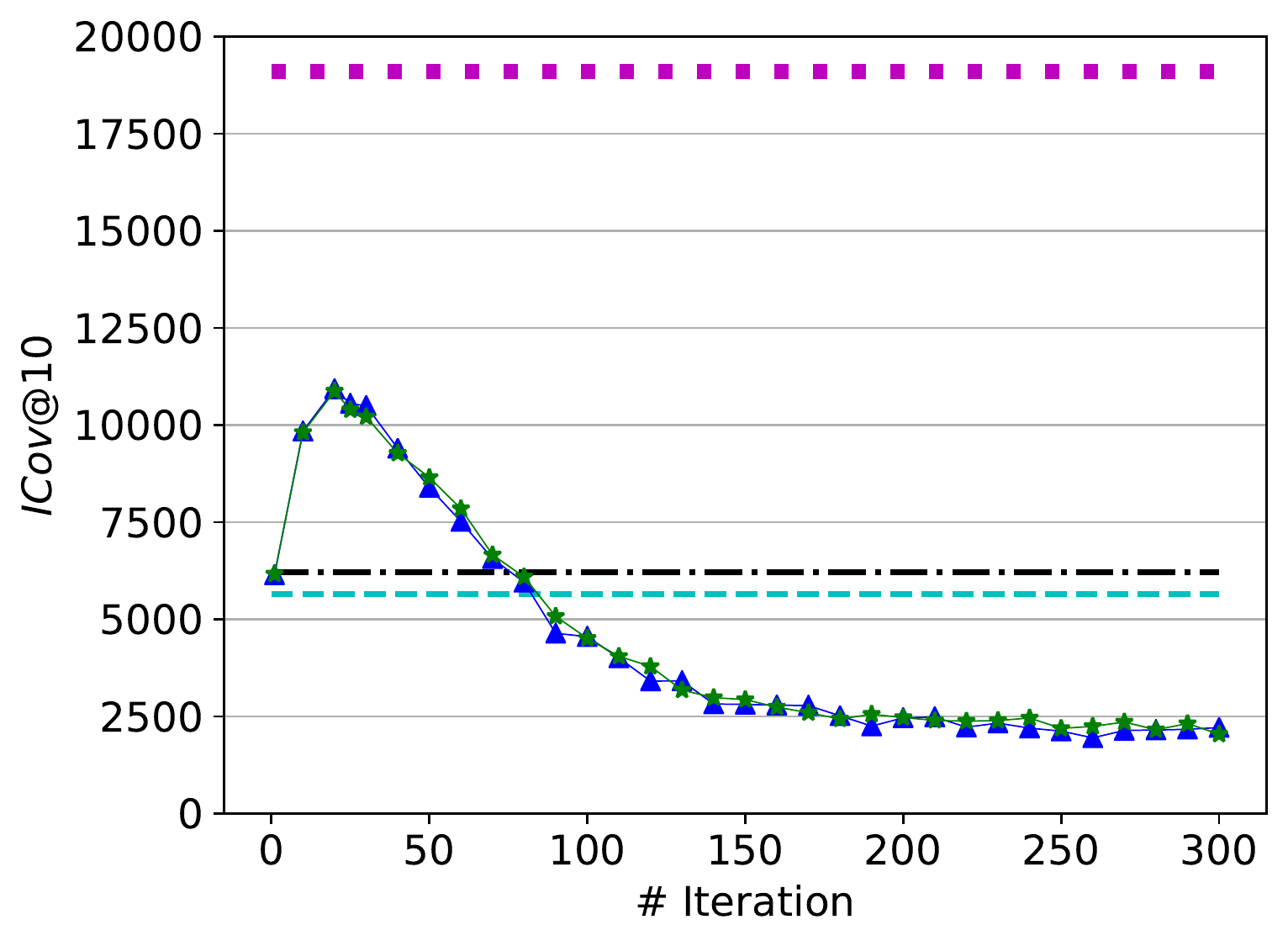}
            \caption{\small Defense-free \textbf{BPR-MF} model.}
            \label{fig:c}
        \end{subfigure}
        \begin{subfigure}[b]{0.45\textwidth}   
            \centering 
            \includegraphics[width=\textwidth]{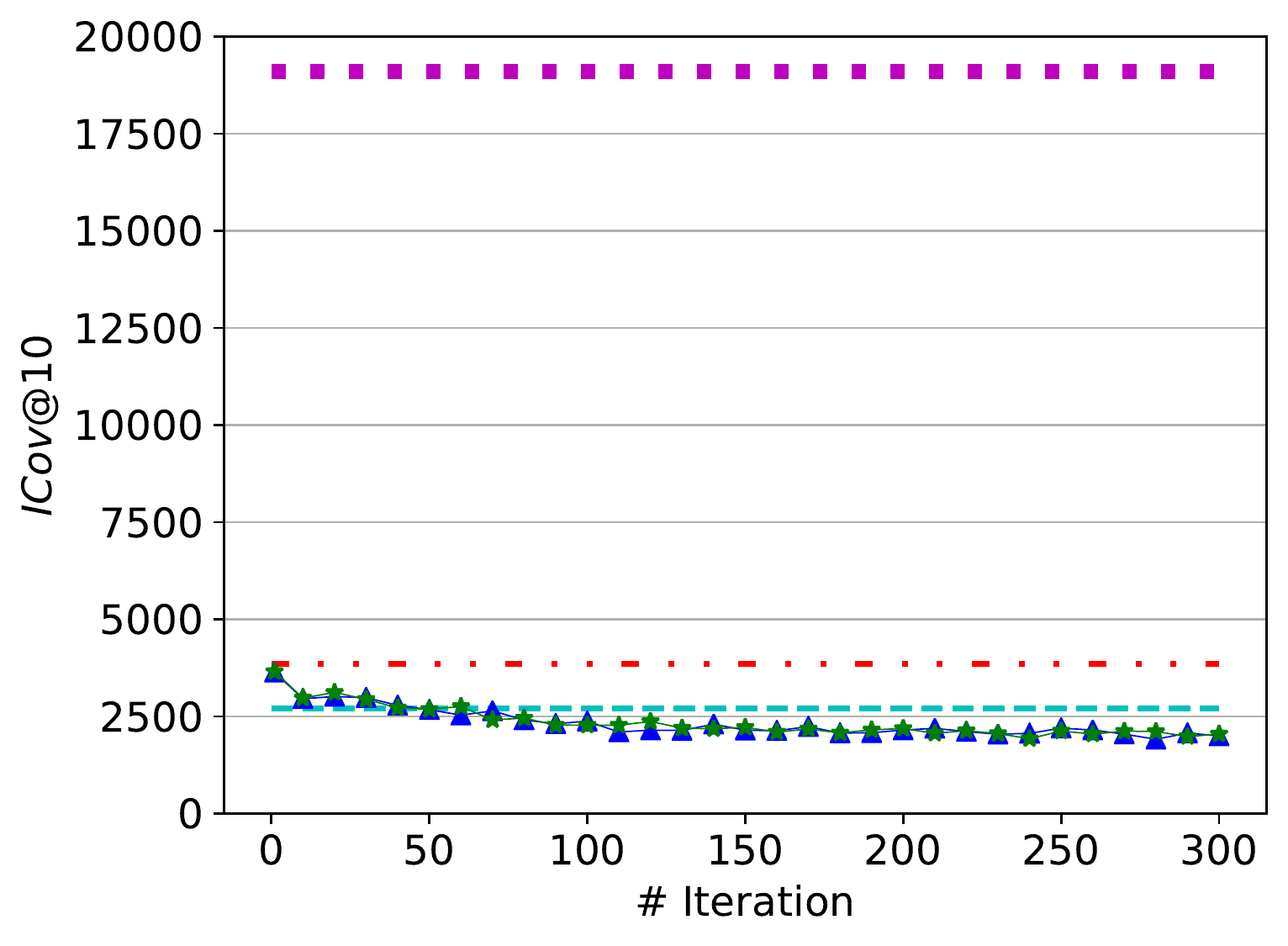}
            \caption{\small Defended BPR-MF model - \textbf{AMF}.} 
            \label{fig:d}
        \end{subfigure}
        \end{adjustbox}
                \vskip\baselineskip
        \begin{subfigure}[b]{\textwidth}  
            \centering
            \includegraphics[width=0.85\textwidth]{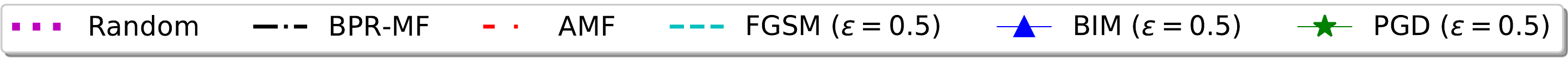}
        \end{subfigure}
        \caption{Results of accuracy ($nDCG$) and diversity ($ICov$) metrics for the \lfm dataset. Figures (a) and (b) show the effects of adversarial perturbations on the $nDCG$ on BPR-MF and AMF respectively. Figures (c) and (d) report the variation of the $ICov$. Figures report also the results of the random recommender (violet dotted line).  }

        \label{fig:overall_effect}
        
\end{figure*}


%% file: 4Discussion.tex
In this section, we discuss the experimental results to answer the previously defined open questions on the effectiveness of iterative perturbations\footnote{Throughout this work, the terms iterative perturbations and multi-step perturbations are used in an interchangeable manner.} and the impact of structural dataset characteristics on the robustness of the model under perturbations. 

\subsection{Investigating the effect of iterative perturbations against the recommender performance accuracy.}

This study aims to measure the effects of adversarial perturbations on accuracy and beyond-accuracy objectives. In particular, to better understand the merits of the presented adversarial perturbations, we aim to answer the following evaluation questions:

\begin{itemize}
    \item \textbf{On the perturbation side}: how much adversarial perturbations obtained from the single-step FGSM and the two iterative BIM and PGD perturbation methods can impair the quality of the original BPR-MF model? The answer to this question can be found in Fig.~\ref{fig:a} \&  Fig.~\ref{fig:c} by comparing the original non-perturbed model with the rest.

    \item \textbf{On the defensive side}: what is the impact of adversarial training on BPR-MF using the state-of-the-art AMF recommender model on the previous question?
    The answer to this question can be found in Fig.~\ref{fig:b} \& Fig.~\ref{fig:d}.
\end{itemize}

In this line, since the performance of the two presented multi-step adversarial perturbations varies based on the number of iterations, in Section~\ref{subsubsec:impact-iterations} we discuss and analyze the effectiveness of the presented perturbations across different
iterations. Afterward, in Section~\ref{subsubsec:impact-epsilon}, we relax the iteration dimension, and for a fixed iteration number, we study how adversarial perturbation budget $\epsilon$ impairs recommendation outcomes of two categories of perturbations (iterative v.s. single-step FGSM).

\input{tables/AUC-ml1m}

\subsubsection{Impact of iterative perturbations based on number of iterations.}~\label{subsubsec:impact-iterations}
Fig.~\ref{fig:overall_effect} visualizes the performance (accuracy and beyond-accuracy) variations by increasing the number of iterations for the \lfm dataset. We use the plot visualization only for the \lfm dataset to show the obtained insights vividly. We show the performance of random recommender to serve as a lower baseline. 
Then, Table~\ref{table:auc} summarizes the complete results obtained across various evaluation metrics and recommendation/perturbation models on both the \ml and \lfm datasets. 
For space limitation, Table~\ref{table:auc} reports the normalized accumulated values of the measured recommendation performance across iterations due to the large set of examined metrics.

On the perturbation side, by looking at Fig~\ref{fig:a}, one can note that both multi-step perturbations, i.e., BIM and PGD, are more powerful compared with the state-of-the-art single-step FGSM perturbation, for a fixed perturbation budget $\epsilon=0.5$. We can realize that only after 25 iterations, the $nDCG$ of PGD drops from the original 0.2033 to 0.0080, which is much lower than FGSM (0.1216).
For instance, the PGD perturbation shows \textbf{15.1} (0.1216 v.s. 0.0080), \textbf{20.4} (0.1216 v.s. 0.0060), and \textbf{23.8} (0.1216 v.s. 0.0051) times stronger impact with respect to FGSM, for iterations 25, 40, and 50 respectively. 
These results confirm the findings in the CV field~\cite{DBLP:conf/iclr/KurakinGB17a} on the superiority of iterative adversarial strategies --- in terms of perturbations' effectiveness --- compared to single-step ones for the item recommendation task. They also demonstrate the vulnerability of the state-of-the-art pairwise ranking model (BPR-MF) against modern adversarial perturbations.
To show the impact better, we can compare the quality of BPR-MF with the random recommender in Fig~\ref{fig:a} in which we can realize after about 25 iterations, the perturbed BPR-MF starts to perform similar to the random recommender. In other words, after few iterations under iterative perturbations, the BPR-MF model loses all the learned information about the individual users' preferences.

The same pattern of results can be noticed by looking at the overall results provided in Table~\ref{table:auc}, where for all <dataset, recommender> combinations, the iterative perturbation strategies (PGD and BIM) outperform FGSM. For instance, the <\ml, BPR-MF> combination shows that the accuracy performance under the PGD perturbation is reduced by more than 2 times compared to FGSM, e.g., (0.0074 v.s. 0.0035), (0.0740 v.s. 0.0353), and (0.0368 v.s. 0.0172) for $PR$, $RE$, and $nDCG$, respectively. Here, we should point out that both Figure~\ref{fig:overall_effect} and Table~\ref{table:auc} do not show a clear difference/advantages of PGD perturbation compared to BIM perturbation. This finding is different from the one previously reported by Athalye \textit{et al.}~\cite{DBLP:conf/icml/AthalyeC018} in CV. We motivate it by the fact that the item recommendation model BPR is less sensitive to the initial value of weights compared with a classification task by using a deep neural network, since in BPR the gradients are computed based on the differences between pairs. 

For what concerns beyond-accuracy analysis, we found an interesting behavior for the defense-free BPR-MF. We can see that during the first 25 iterations of BIM, $ICov$ increments nearly by 76\% (from 6,220 to 10,928) compared to the coverage value of the non-perturbed recommender (see Fig.~\ref{fig:c}). After that, it steadily diminishes with a minimum $ICov$ value of 1,948 (for BIM). 
This result may be justified by the fact that when the multi-step perturbation has computed several iterations ($L \geq 70$), it steadily destructs the accuracy metrics, and brings the model to recommend a set of few items that all the users will not appreciate. Thus, we can note that the proposed iterative perturbations impair the personalized recommender to perform as bad as a random recommender (see Fig.~\ref{fig:a}) concerning accuracy metric and even worse with respect to beyond-accuracy metrics. Table~\ref{table:auc} confirms low beyond-accuracy metric values ($EFD$ and the $SE$) for BIM and PGD perturbations, with reduction factor equal to or more than $50\%$ and $20\%$ respectively, independently from the <dataset, recommender, iterative-perturbation> combination.

On the defensive side, we studied the impact of adversarial training, known as APR, on various perturbation strategies. We observed an evident performance drop in accuracy for AMF (see Figure~\ref{fig:b}), which is, on average, more than 58\% for multi-step strategies and 11.31\% for the single-step one (see Table~\ref{table:auc}). For instance, the PGD perturbation shows \textbf{1.48} (0.2147 v.s. 0.1448), \textbf{1.86} (0.2147 v.s. 0.1154), and \textbf{1.94} (0.2147 v.s.  0.1106) times stronger impact with respect to FGSM, for iterations 20, 30, and 50 respectively. However, the accuracy reduction does not reach random performance as for the BPR-MF recommender. We may explain this slight decrease/increase in robustness by mentioning the partial effectiveness of the adversarial regularization procedure (i.e., specifically designed to protect against single-step perturbations~\cite{DBLP:conf/sigir/0001HDC18}).

\subsubsection{Impact of iterative perturbations based on $\epsilon$-limited budget perturbation.}~\label{subsubsec:impact-epsilon}
\input{images/figures-eps} 

In this study, we relax the investigation of the impact of iteration increase on iterative perturbations' performances. Instead, by fixing the number of iterations (i.e., $L=25$, the value previously shown to be the critical point (the elbow of the curve in Fig.~\ref{fig:a}) in performance deterioration) and varying $\epsilon$ from $0.001$ to $10$, we investigate at what  $\epsilon$-level, iterative perturbations can get a similar performance comparable with FGSM. Analyzing Fig.~\ref{fig:eps-a} \&~\ref{fig:eps-b}, we found that iterative adversarial strategies reach the FGSM ($\epsilon=0.5$) performance at iteration-level $\epsilon \simeq 0.1$. In other words, by using $\nicefrac{0.5}{0.1}=$\textbf{5} times less perturbation budget, the new iterative strategies reach a similar performance as that of the state-of-the-art FGSM perturbation strategy, independently from the recommender, i.e., the defense-free BPR-MF or the adversarial defensed AMF.

In summary, the results of two above studies provides strong evidence that:
\begin{itemize}
    \item \textit{\textbf{iterative perturbations} for the item recommendation task are more powerful than the single-step strategies widely adopted in the prior literature of RS community. For example, let the perturbation budget fixed to $\epsilon = 0.5$, the multi-step strategies reduce the BPR-MF performance by an amount of 15 times (along $nDCG$) with only 25 iterations, while, let the caused performance degradation fixed, iterative strategies are as effective as single-step ones by using only the 20\% of the perturbation budget ($\epsilon \simeq 0.1$);}     
    \item  \textit{the state-of-the-art \textbf{robustification} strategy adopted in the RS community, the adversarial regularization, can diminish the impact of iterative perturbations. However, the iterative perturbations still have a high capability to impact and impair the quality of the defended recommender. These results suggest the need to identify mediating factors that can reduce the impact of iterative perturbations against RS, but it is left for future investigation.}
\end{itemize}

\subsection{Studying the impact of structural dataset characteristics.}
\input{images/figure-sparsity}

The second part of our analysis relates to the effect of structural dataset characteristics on adversarial perturbations' effectiveness. 
Inspired by the exploratory study on the impact of dataset characteristics on the accuracy~\cite{DBLP:journals/tmis/AdomaviciusZ12} and shilling perturbations robustness~\cite{sisinflabSigir2020} of RSs, we evaluated the effect of the \textit{density}, \textit{size}, and \textit{shape} structural features in the variation of an accuracy metric ($nDCG$) under adversarial perturbation on the model parameters, defined by $ \nicefrac{|\mathcal{F}|}{|\mathcal{U}| \times |\mathcal{I}|}$, $|\mathcal{U}| \times |\mathcal{I}|$, and $\nicefrac{|\mathcal{U}|}{|\mathcal{I}|}$ respectively. 
We produced 10 sub-samples for each dataset with different structural properties. Each sub-sample is a \textit{k-core} version of the original datasets, such that we removed users and items with less than \textit{k} recorded interactions.
Note that increasing the \textit{k} from 10 to 100, enabled us to generate sub-samples with increasing levels of \textit{density}. 

Figure~\ref{fig:impact-data-char} shows the relative percentage variation of the accuracy metric with respect to the reference $nDCG_{init}$ (performance before the perturbation), defined by $\rho = \frac{\Delta nDCG}{nDCG_{init}} \times 100$ where $\Delta nDCG = nDCG_{init} - nDCG_{after}$ represents the change on $nDCG$ before and after the adversarial perturbation.
In line with what has been shown in~\cite{sisinflabSigir2020} on the impact of dataset characteristics on the effectiveness of hand-engineered shilling perturbations against CF models, our results as well confirm the relationship between the \textit{density} of the dataset and \textit{adversarial perturbations' effectiveness}. This means, increasing density of the dataset tends to reduce the impact of perturbation measured by $\rho$. For instance, the PGD perturbation against <\ml, BPR-MF> combination with sub-samples having \textit{densities} equal to 0.0507 and 0.1387, caused a performance reduction of 43.79\%, and 28.90\%. Thus, we can observe that denser (i.e., less sparse) datasets protect recommender models against adversarial perturbations. This can be explained by the fact that increasing density inherently increases inter-dependency among users and items and can improve robustness, as also mentioned by Christakopoulou et al.~\cite{DBLP:conf/recsys/Christakopoulou19}.

For what concerns investigation on the impact of \textit{size} and the \textit{shape} characteristics, we report in Figures.~\ref{fig:size-b} to~\ref{fig:shape-f}, the values obtained for $\rho$ by increasing the \textit{size} and \textit{shape} of the datasets (sub-samples). We found that these characteristics show similar behavior, i.e., increasing shape/size implies increments $\rho$ (impact of perturbations), and these results are consistent with findings for shilling perturbation analysis in~\cite{sisinflabSigir2020}. 

Finally, we examined the impact of defense strategy on the previous insights about the effect of data characteristics. Fig.~\ref{fig:density-a} \&~\ref{fig:density-d} show that the influence of \textit{density} on $\rho$ is much higher in BPR-MF than in AMF. For example, the slope\footnote{Slope has been calculated using the formula $m=\frac{y_2-y_1}{x_2-x_1}$ where $(x_1, y_1)$ and $(x_2, y_2)$ represent two points in the line.} of the PGD fitted line evaluated is -96.10 and -3.65 for BPR-MF and AMF respectively (note that scales in the y-axis for plots shown for Fig.~\ref{fig:density-a} and \ref{fig:density-d} are different). The same pattern is visible for the \textit{shape} and \textit{size} properties. We can conclude that dataset characteristics do not influence the efficacy of adversarial perturbations against adversarial defended models as much as in the defense-free setting.  

We can summarize the findings as follows: \textit{structural dataset characteristics, i.e., shape, space, and density, impact the effectiveness of adversarial perturbations against model-based CF recommenders such that adversarial perturbations are less effective by either increasing the density or decreasing the shape/size of the dataset. Furthermore, the analysis on the impact of dataset characteristics confirms that APR, applied on AMF, reduces the effectiveness of both single and multi-step adversarial perturbations.}

%% file: tables/AUC-ml1m.tex
\begin{table*}[!t]
\caption{Comparison of Initial not-perturbed performance and single-step FGSM with the accumulated normalized values of the accuracy and beyond-accuracy metrics across the multi-step adversarial perturbation strategies (see Figure~\ref{fig:overall_effect}) evaluated on the top-\textbf{10} recommendation lists. The perturbation budget $\epsilon$ is $0.5$. We put in \textbf{bold} the lower value (the perturbation is more effective).}

\begin{tabular}{ll|cccc|cccc}
\toprule

\multicolumn{1}{c}{\multirow{2}{*}{\textbf{Model}}} & \multicolumn{1}{c}{\multirow{2}{*}{\begin{tabular}[c]{@{}c@{}}\textbf{Metric}\\\textbf{@10}\end{tabular}}}& \multicolumn{ 4}{c}{\lfm} & \multicolumn{ 4}{c}{\ml} \\ 
 \cmidrule(lr){3-6} \cmidrule(lr){7-10}

\multicolumn{ 1}{l}{} & \multicolumn{ 1}{c}{} & \multicolumn{1}{c}{Initial} & \multicolumn{1}{c}{FGSM} & \multicolumn{1}{c}{BIM} & \multicolumn{1}{c}{PGD} & \multicolumn{1}{c}{Initial} & \multicolumn{1}{c}{FGSM} & \multicolumn{1}{c}{BIM} & \multicolumn{1}{c}{PGD} \\ \midrule
\multicolumn{1}{c}{\multirow{6}{*}{\textbf{BPR-MF}}} 

& $PR$ & 0.0310 & 0.0211 & 0.0019 & \textbf{0.0018} & 0.0088 & 0.0074 &\textbf{ 0.0035} & \textbf{0.0035} \\ 
& $RE$ & 0.3102 & 0.2115 & 0.0194 & \textbf{0.0177} & 0.0884 & 0.0740 & \textbf{0.0353} & \textbf{0.0353} \\ 
& $nDCG$ & 0.2033 & 0.1216 & 0.0111 & \textbf{0.0100} & 0.0447 & 0.0368 & 0.0174 & \textbf{0.0172} \\ 
& $EFD$ & 0.5144 & 0.3069 & 0.0313 & \textbf{0.0284} & 0.0977 & 0.0791 & 0.0355 & \textbf{0.0353} \\ 
& $SE$ & 11.3454 & 11.1359 & \textbf{10.1703} & 10.2141 & 9.6306 & 9.1582 & \textbf{7.4000} & 7.4457 \\ 
& $ICov$ & 6220 & 5645 & \textbf{4352} & 4428 &   2247 & 2433 & \textbf{1189} & 1213 \\

\midrule
\multicolumn{1}{c}{\multirow{6}{*}{\textbf{AMF}}}  

& $PR$ & 0.0357 & 0.0316 & \textbf{0.0164} & 0.0167 & 0.0092 & 0.0085 & \textbf{0.0048} & \textbf{0.0048} \\ 
& $RE$ & 0.3565 & 0.3165 & \textbf{0.1644} & 0.1667 & 0.0922 & 0.0846 & \textbf{0.0482} & 0.0484 \\ 
& $nDCG$ & 0.2421 & 0.2147 & \textbf{0.1010} & 0.1030 & 0.0462 & 0.0419 & \textbf{0.0228} & 0.0231 \\ 
& $EFD$ & 0.5987 & 0.5184 & \textbf{0.2303} & 0.2352 & 0.0971 & 0.0853 & \textbf{0.0442} & 0.0447 \\ 
& $SE$ & 9.9758 & 8.8980 & \textbf{7.1871} & 7.1993 & 8.3049 & 7.4123 & \textbf{6.2958} & 6.2984 \\ 
& $ICov$ & 3847  & 2708 & \textbf{2315} & 2321 & 1486   & 1169 & \textbf{1066} & 1077  \\

\bottomrule
\end{tabular}
\label{table:auc}
\end{table*}

%% file: images/figures-eps.tex
\begin{figure}[]
        \centering
        \begin{adjustbox}{minipage=\linewidth}
        \begin{subfigure}[b]{0.47\textwidth}   
            \centering 
            \includegraphics[width=\textwidth]{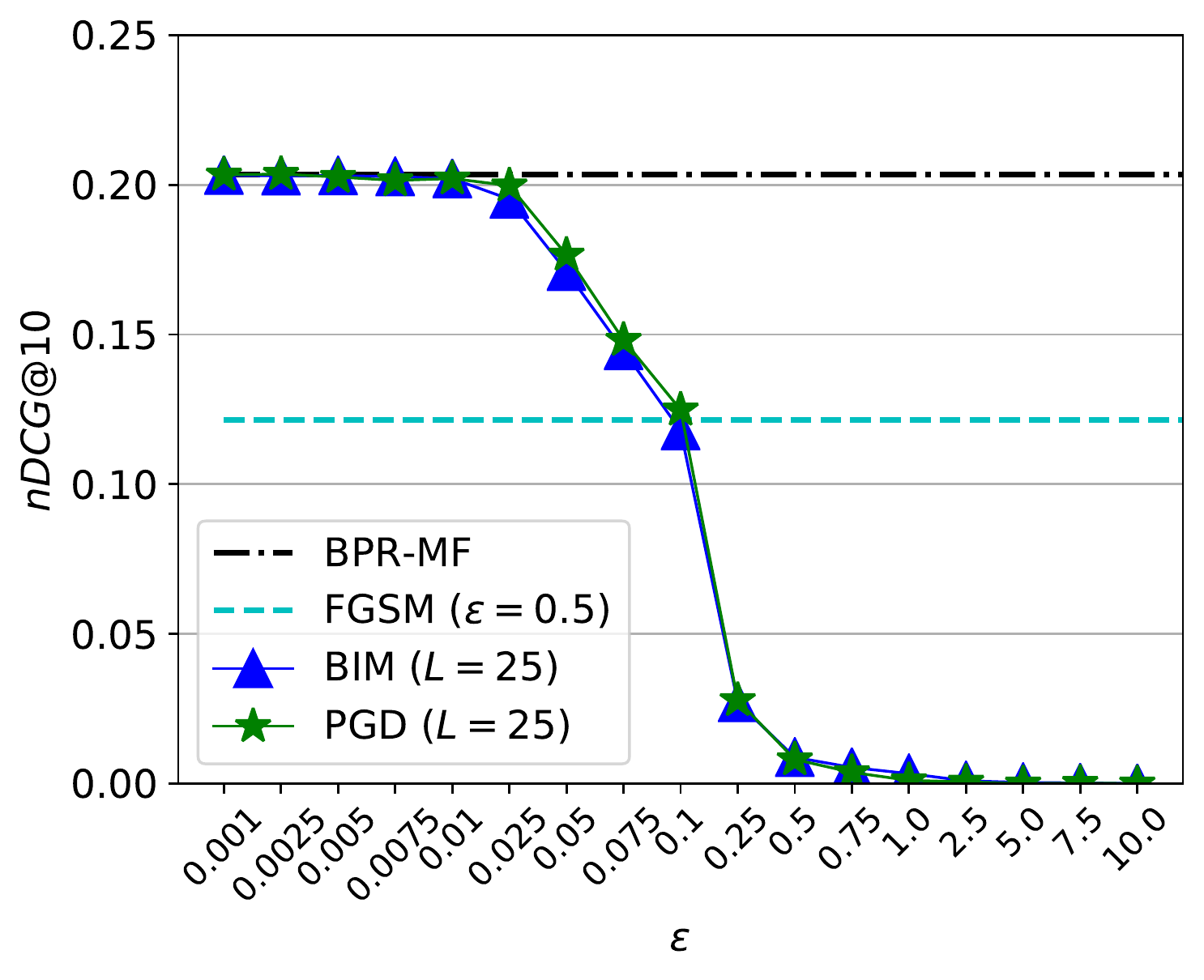}
            \caption{\small \textbf{BPR-MF} (varying $\epsilon$)}  
            \label{fig:eps-a}
        \end{subfigure}
        \quad
        \begin{subfigure}[b]{0.47\textwidth}
            \centering
            \includegraphics[width=\textwidth]{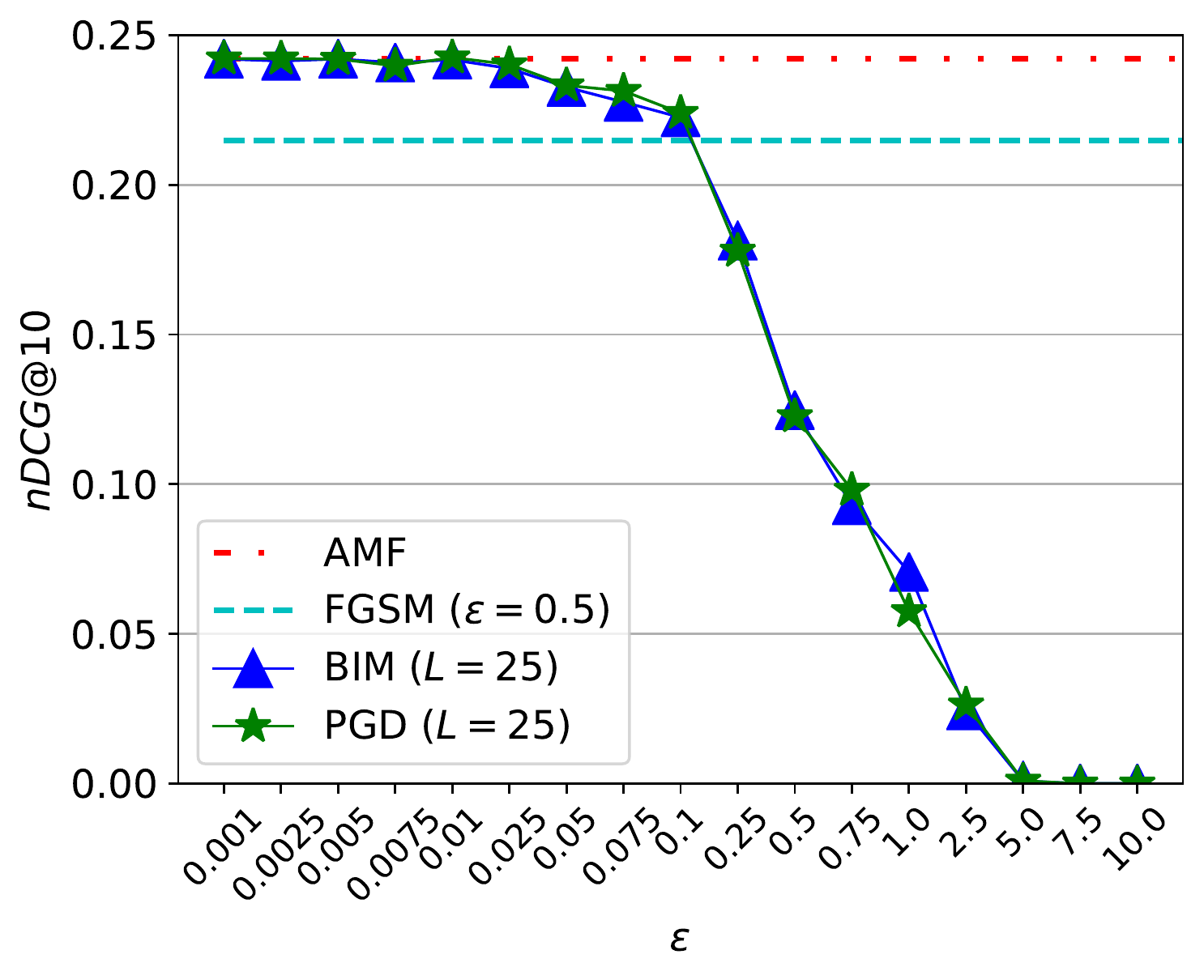} \caption{\small \textbf{AMF} (varying $\epsilon$)}    
            \label{fig:eps-b}
        \end{subfigure}
     \end{adjustbox}
    \caption{\small Results on iterative perturbations with a fixed number of iterations $L=25$ and increasing budget perturbation values $\epsilon \in [0.001, 10.0]$ on \lfm dataset. Figure~\ref{fig:eps-a} and~\ref{fig:eps-b} shows that with a small perturbation e.g., $\epsilon \simeq 0.1$, multi-step noises are more effective than a single-step ones with a bigger perturbation budget ($\epsilon=0.5$). } 
    \label{fig:eps}

\end{figure}

%% file: images/figure-sparsity.tex
\begin{figure*}
        \centering
    \begin{adjustbox}{minipage=\linewidth,scale=0.88}
    
        \begin{subfigure}[b]{0.33\textwidth}   
            \centering 
            \includegraphics[width=\textwidth]{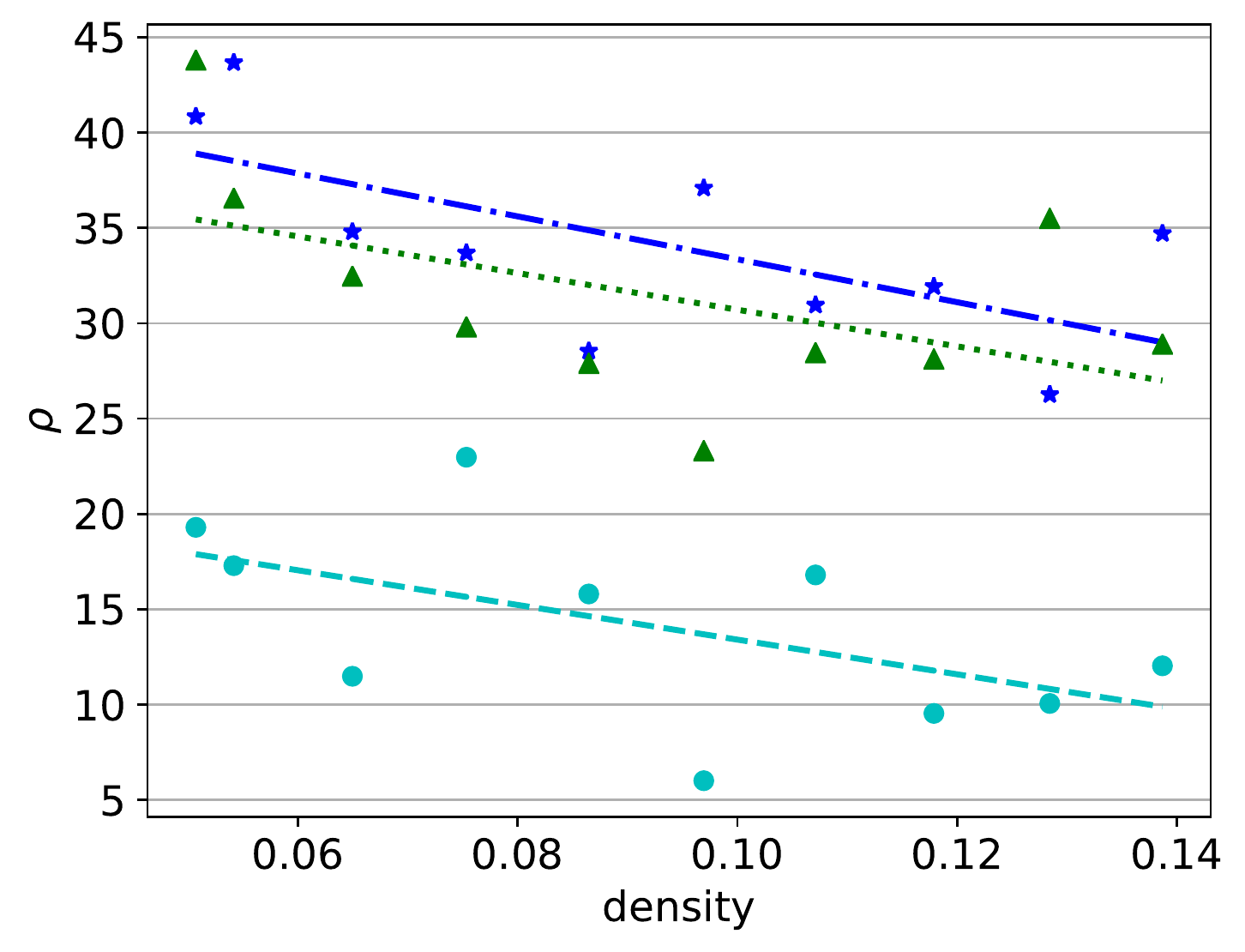}
            \caption{\small \textbf{BPR-MF} (varying \textit{density}.)}    
            \label{fig:density-a}
        \end{subfigure}
        \quad
          \begin{subfigure}[b]{0.33\textwidth}   
                \centering 
                \includegraphics[width=\textwidth]{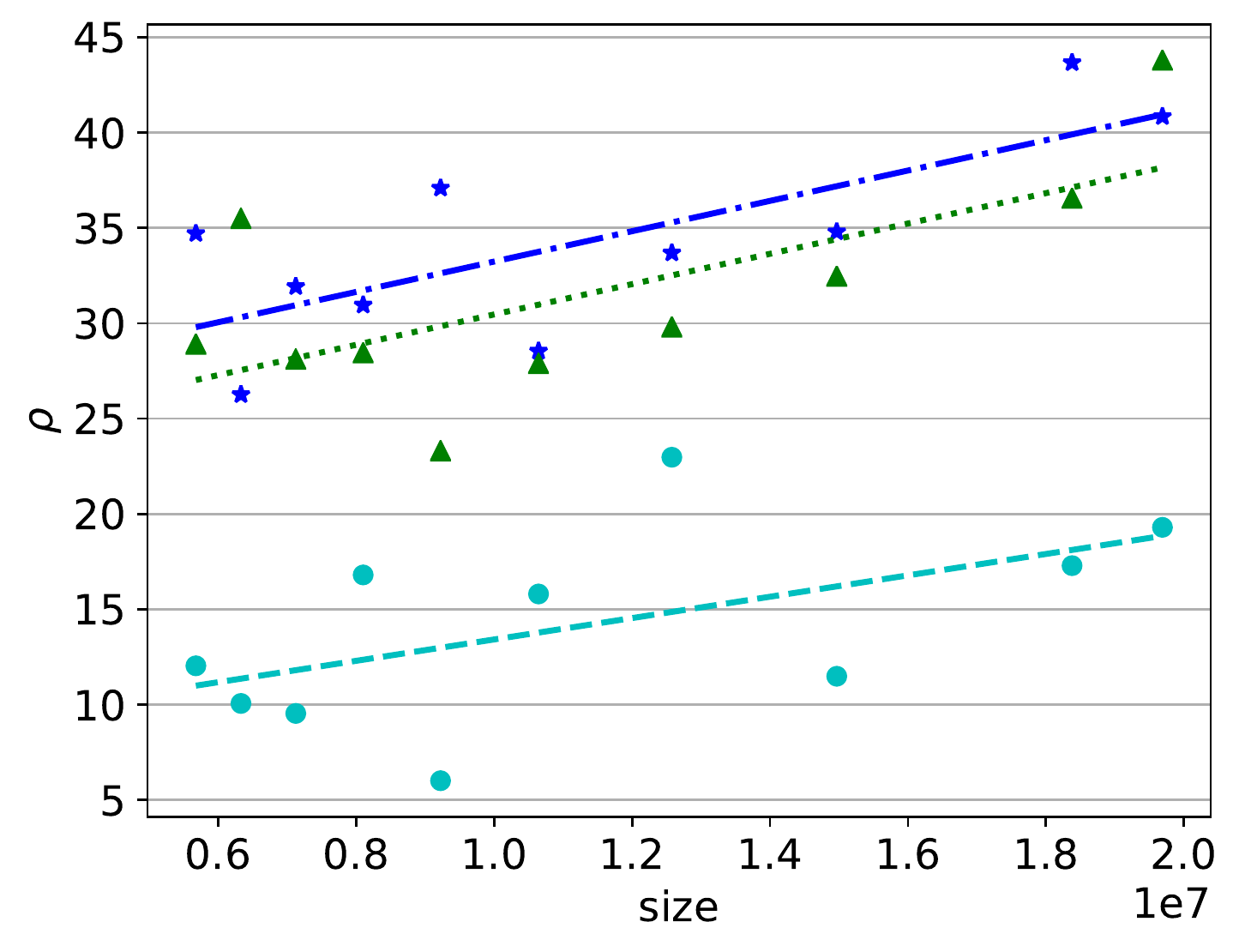}
                \caption{\small \textbf{BPR-MF} (varying \textit{size}.)}   
                \label{fig:size-b}
            \end{subfigure}
        \quad
          \begin{subfigure}[b]{0.33\textwidth}   
                \centering 
                \includegraphics[width=\textwidth]{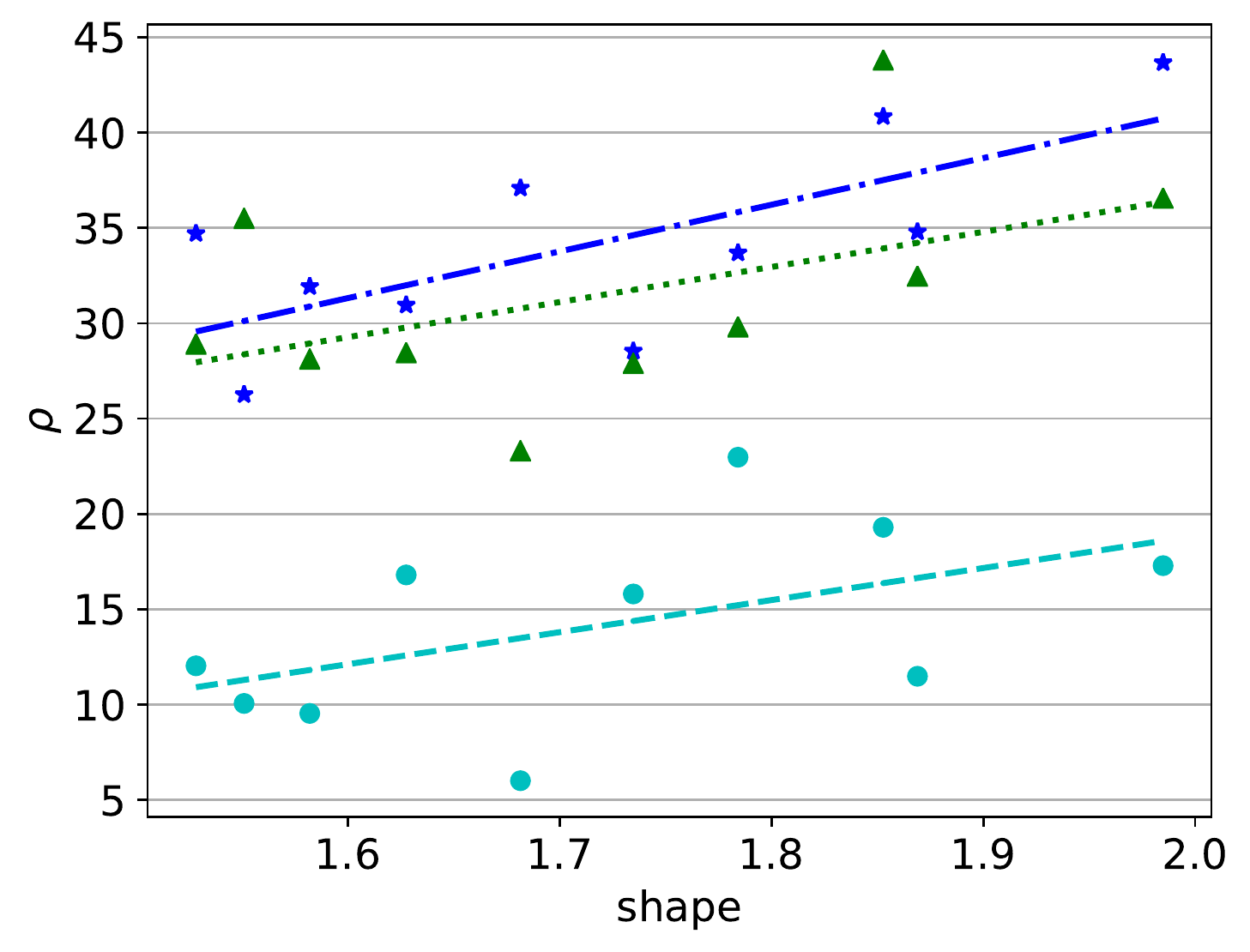}
                \caption{\small \textbf{BPR-MF} (varying \textit{shape}.)}   
                \label{fig:shape-c}
            \end{subfigure}
            
    \vskip\baselineskip

         \begin{subfigure}[b]{0.33\textwidth}
            \centering
            \includegraphics[width=\textwidth]{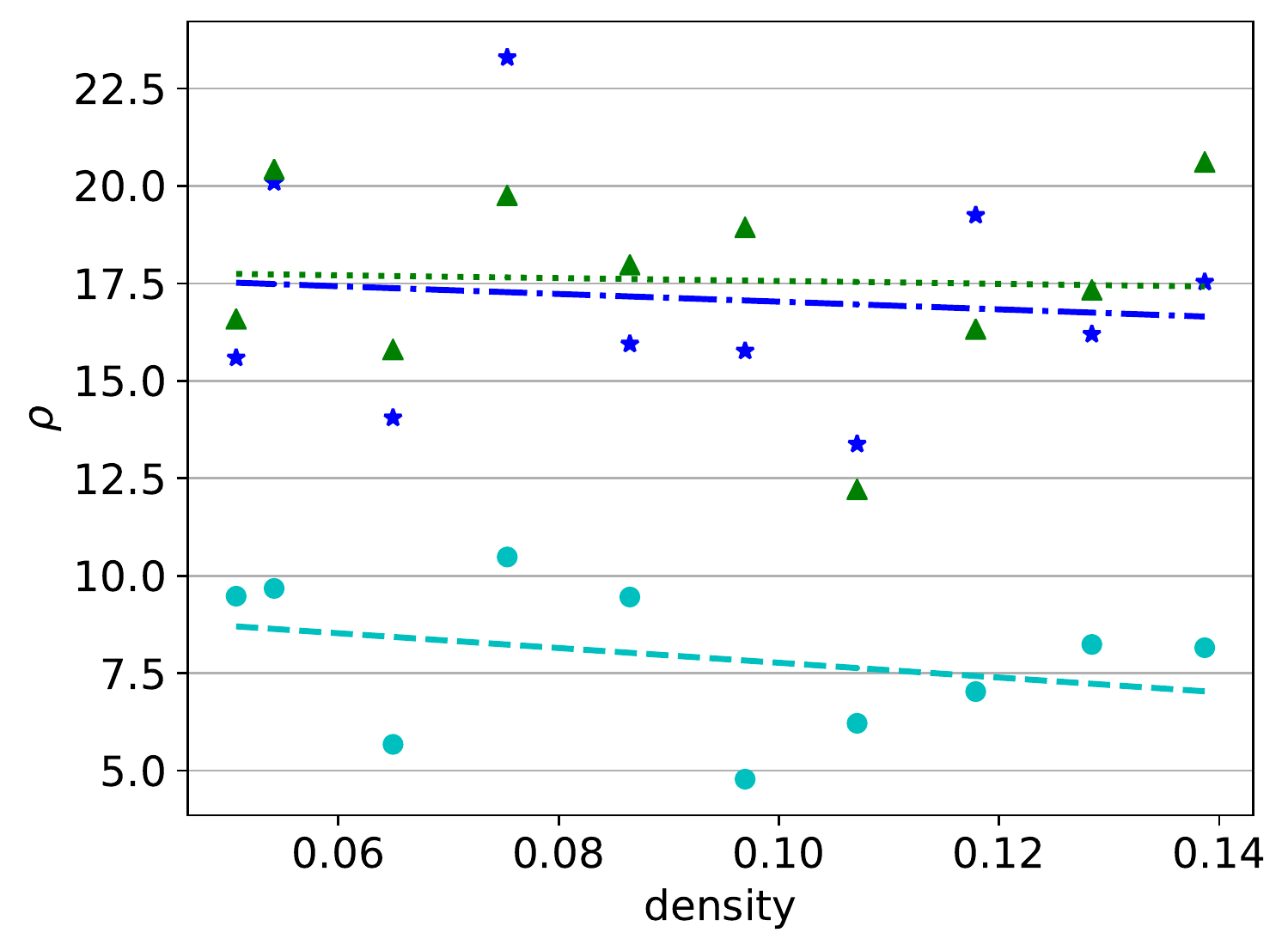}
            \caption{\small \textbf{AMF} (varying \textit{density}.)}   
                \label{fig:density-d}
        \end{subfigure}
        \quad
        \begin{subfigure}[b]{0.33\textwidth}
            \centering
            \includegraphics[width=\textwidth]{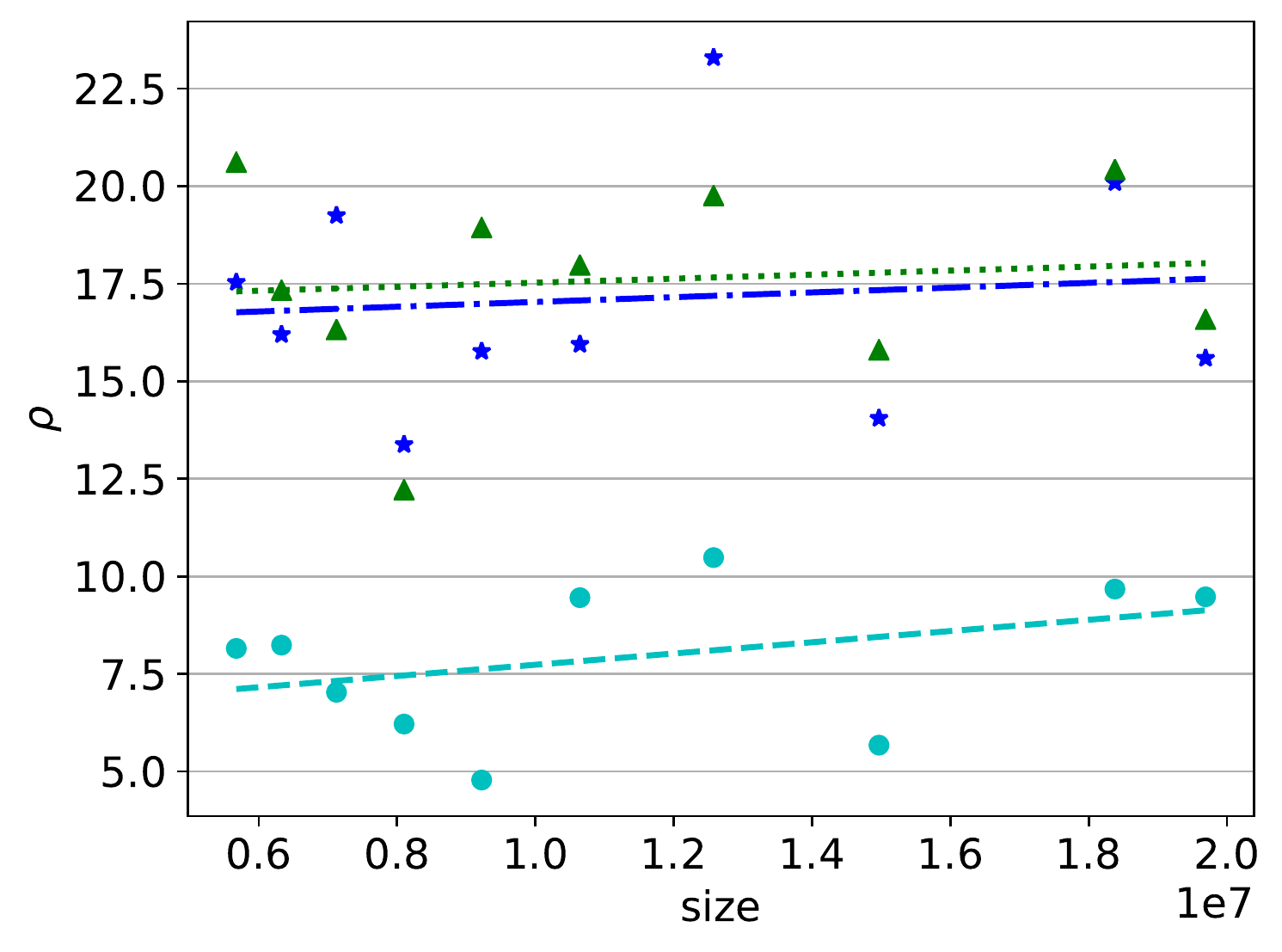}
            \caption{\small \textbf{AMF} (varying \textit{size}.)}   
                \label{fig:size-e}
        \end{subfigure}
                \quad
        \begin{subfigure}[b]{0.33\textwidth}
            \centering
            \includegraphics[width=\textwidth]{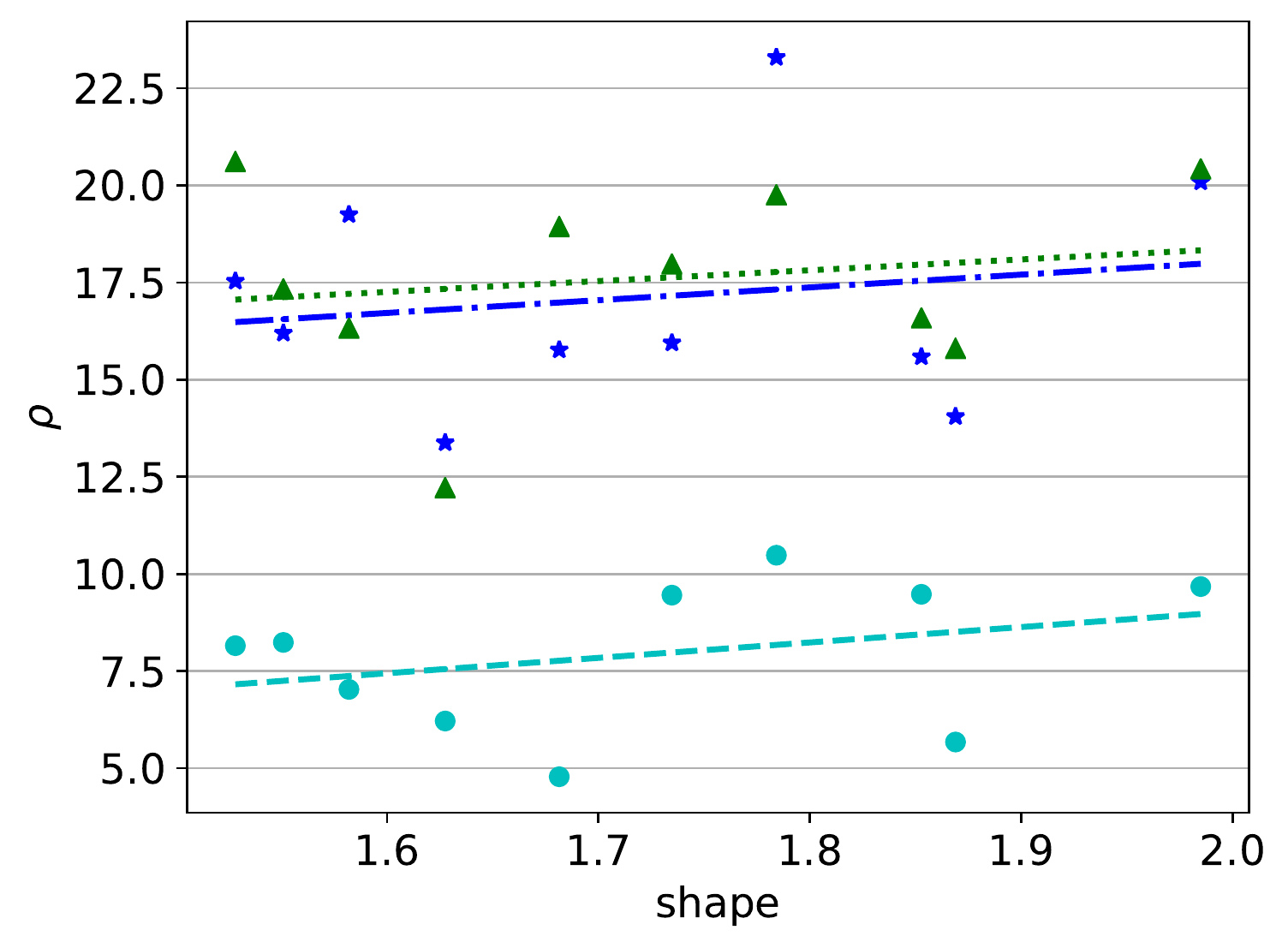}
            \caption{\small \textbf{AMF} (varying \textit{shape}.)} 
                \label{fig:shape-f}
        \end{subfigure}
    \end{adjustbox}
    
        \vskip\baselineskip
        \begin{subfigure}[b]{\textwidth}  
            \centering
            \includegraphics[width=0.60\textwidth]{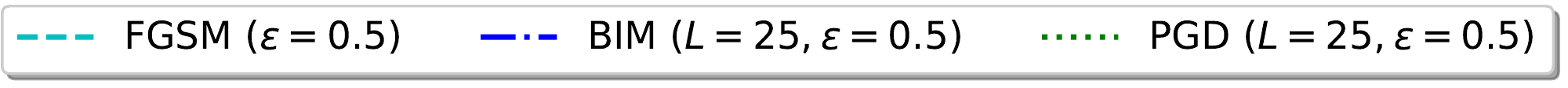}
        \end{subfigure}    
    
    \caption{\small Results of applying the single-step and multi-step iterative adversarial perturbation against \textbf{BPR-MF} (first-row) and \textbf{AMF} (second-row) models trained on different sub-samples of \ml dataset. The $\rho$ measures the relative percentage variation of the $nDCG$ before and after perturbations. Decreasing trends indicates that increasing in the  dataset characteristic values (e.g.,density), the perturbation is less effective. We plot the fitted lines (using least-square cost) to facilitate compassion between models. }
    \label{fig:impact-data-char}

\end{figure*}

%% file: tables/results_fair.tex
\newcommand{\Def}{AMF\xspace}
\newcommand{\noDef}{BPR-MF\xspace}

\begin{table*}[]
\caption{
Performance (measured in terms of nDCG) of the different approaches on each subset of users/items, where $C_1$ and $C_4$ denote the least and most popular items (for item popularity) and users with less and more interactions, respectively; for user gender $C_1$ is associated to males and $C_2$ to females.
Results for \ml are presented on the left, \lfm on the right.
Best results for each model are highlighted in bold.
}
\label{t:q1}
\resizebox{1.\textwidth}{!}{
\input{tables/table.ml.clusters.ndcg.tex}
\quad
\input{tables/table.lastfm.clusters.ndcg.tex}
}
\end{table*}

\begin{table*}[]
\caption{
Fairness measured according to GCE with $f_i$ fair distributions ($f_0$ always represents a uniform distribution, whereas $f_k$ denotes a distribution where group $C_k$ accumulates more probability than the rest, as in $f_1=[0.75, 0.25]$ for user gender), MAD rating (MADr), and MAD ranking (MADR).
Results for \ml on top and \lfm on bottom; rest of notation as in Table~\ref{t:q1}.
}
\label{t:q4}
\resizebox{1.\textwidth}{!}{
\input{tables/table.ml.item_user.fair_usravg.tex}
}
\resizebox{1.\textwidth}{!}{
\input{tables/table.lastfm.item_user.fair_usravg.tex}
}
\end{table*}

%% file: tables/table.ml.clusters.ndcg.tex
\begin {tabular}{rrrrrrrrrrrr}%
\toprule &&\multicolumn {4}{c}{Item pop} & \multicolumn {2}{c}{User gender} & \multicolumn {4}{c}{User interactions} \\Model&&$C_1$&$C_2$&$C_3$&$C_4$&$C_1$&$C_2$&$C_1$&$C_2$&$C_3$&$C_4$\\\cmidrule (r){1-2}\cmidrule (rl){3-6}\cmidrule (rl){7-8}\cmidrule (rl){9-12}%
\multirow {4}{*}{\noDef }&initial&$\bf \pgfutilensuremath {0.054}$&$\bf \pgfutilensuremath {0.035}$&$\bf \pgfutilensuremath {0.045}$&$\bf \pgfutilensuremath {0.300}$&$\bf \pgfutilensuremath {0.046}$&$\bf \pgfutilensuremath {0.043}$&$\bf \pgfutilensuremath {0.079}$&$\bf \pgfutilensuremath {0.044}$&\pgfutilensuremath {0.032}&$\bf \pgfutilensuremath {0.023}$\\%
&FGSM&\pgfutilensuremath {0.027}&\pgfutilensuremath {0.017}&\pgfutilensuremath {0.043}&\pgfutilensuremath {0.284}&\pgfutilensuremath {0.044}&\pgfutilensuremath {0.041}&\pgfutilensuremath {0.073}&\pgfutilensuremath {0.044}&$\bf \pgfutilensuremath {0.032}$&\pgfutilensuremath {0.022}\\%
&BIM&\pgfutilensuremath {0.005}&\pgfutilensuremath {0.000}&\pgfutilensuremath {0.000}&\pgfutilensuremath {0.167}&\pgfutilensuremath {0.019}&\pgfutilensuremath {0.016}&\pgfutilensuremath {0.018}&\pgfutilensuremath {0.020}&\pgfutilensuremath {0.018}&\pgfutilensuremath {0.016}\\%
&PGD&\pgfutilensuremath {0.000}&\pgfutilensuremath {0.000}&\pgfutilensuremath {0.000}&\pgfutilensuremath {0.178}&\pgfutilensuremath {0.017}&\pgfutilensuremath {0.016}&\pgfutilensuremath {0.022}&\pgfutilensuremath {0.018}&\pgfutilensuremath {0.015}&\pgfutilensuremath {0.012}\\%
\cmidrule (r){1-2}\cmidrule (rl){3-6}\cmidrule (rl){7-8}\cmidrule (rl){9-12}\multirow {4}{*}{\Def }&initial&$\bf \pgfutilensuremath {0.172}$&\pgfutilensuremath {0.096}&\pgfutilensuremath {0.096}&$\bf \pgfutilensuremath {0.334}$&$\bf \pgfutilensuremath {0.047}$&$\bf \pgfutilensuremath {0.043}$&$\bf \pgfutilensuremath {0.078}$&$\bf \pgfutilensuremath {0.047}$&$\bf \pgfutilensuremath {0.034}$&$\bf \pgfutilensuremath {0.026}$\\%
&FGSM&\pgfutilensuremath {0.163}&$\bf \pgfutilensuremath {0.114}$&$\bf \pgfutilensuremath {0.110}$&\pgfutilensuremath {0.326}&\pgfutilensuremath {0.043}&\pgfutilensuremath {0.039}&\pgfutilensuremath {0.070}&\pgfutilensuremath {0.041}&\pgfutilensuremath {0.033}&\pgfutilensuremath {0.022}\\%
&BIM&\pgfutilensuremath {0.000}&\pgfutilensuremath {0.000}&\pgfutilensuremath {0.000}&\pgfutilensuremath {0.198}&\pgfutilensuremath {0.022}&\pgfutilensuremath {0.018}&\pgfutilensuremath {0.024}&\pgfutilensuremath {0.018}&\pgfutilensuremath {0.025}&\pgfutilensuremath {0.018}\\%
&PGD&\pgfutilensuremath {0.002}&\pgfutilensuremath {0.055}&\pgfutilensuremath {0.000}&\pgfutilensuremath {0.202}&\pgfutilensuremath {0.023}&\pgfutilensuremath {0.017}&\pgfutilensuremath {0.024}&\pgfutilensuremath {0.018}&\pgfutilensuremath {0.025}&\pgfutilensuremath {0.018}\\\bottomrule %
\end {tabular}%

%% file: tables/table.lastfm.clusters.ndcg.tex
\begin {tabular}{rrrrrrrrrrrr}%
\toprule &&\multicolumn {4}{c}{Item pop} & \multicolumn {2}{c}{User gender} & \multicolumn {4}{c}{User interactions} \\Model&&$C_1$&$C_2$&$C_3$&$C_4$&$C_1$&$C_2$&$C_1$&$C_2$&$C_3$&$C_4$\\\cmidrule (r){1-2}\cmidrule (rl){3-6}\cmidrule (rl){7-8}\cmidrule (rl){9-12}%
\multirow {4}{*}{\noDef }&initial&$\bf \pgfutilensuremath {0.000}$&\pgfutilensuremath {0.000}&$\bf \pgfutilensuremath {0.006}$&$\bf \pgfutilensuremath {0.092}$&$\bf \pgfutilensuremath {0.218}$&$\bf \pgfutilensuremath {0.143}$&$\bf \pgfutilensuremath {0.158}$&$\bf \pgfutilensuremath {0.209}$&$\bf \pgfutilensuremath {0.194}$&$\bf \pgfutilensuremath {0.253}$\\%
&FGSM&\pgfutilensuremath {0.000}&$\bf \pgfutilensuremath {0.001}$&\pgfutilensuremath {0.004}&\pgfutilensuremath {0.062}&\pgfutilensuremath {0.131}&\pgfutilensuremath {0.085}&\pgfutilensuremath {0.102}&\pgfutilensuremath {0.118}&\pgfutilensuremath {0.123}&\pgfutilensuremath {0.143}\\%
&BIM&\pgfutilensuremath {0.000}&\pgfutilensuremath {0.000}&\pgfutilensuremath {0.000}&\pgfutilensuremath {0.004}&\pgfutilensuremath {0.007}&\pgfutilensuremath {0.009}&\pgfutilensuremath {0.011}&\pgfutilensuremath {0.007}&\pgfutilensuremath {0.009}&\pgfutilensuremath {0.002}\\%
&PGD&\pgfutilensuremath {0.000}&\pgfutilensuremath {0.001}&\pgfutilensuremath {0.000}&\pgfutilensuremath {0.002}&\pgfutilensuremath {0.004}&\pgfutilensuremath {0.006}&\pgfutilensuremath {0.007}&\pgfutilensuremath {0.005}&\pgfutilensuremath {0.004}&\pgfutilensuremath {0.004}\\%
\cmidrule (r){1-2}\cmidrule (rl){3-6}\cmidrule (rl){7-8}\cmidrule (rl){9-12}\multirow {4}{*}{\Def }&initial&\pgfutilensuremath {0.000}&$\bf \pgfutilensuremath {0.006}$&$\bf \pgfutilensuremath {0.014}$&$\bf \pgfutilensuremath {0.106}$&$\bf \pgfutilensuremath {0.260}$&$\bf \pgfutilensuremath {0.188}$&$\bf \pgfutilensuremath {0.174}$&$\bf \pgfutilensuremath {0.237}$&$\bf \pgfutilensuremath {0.229}$&$\bf \pgfutilensuremath {0.329}$\\%
&FGSM&\pgfutilensuremath {0.000}&\pgfutilensuremath {0.000}&\pgfutilensuremath {0.010}&\pgfutilensuremath {0.095}&\pgfutilensuremath {0.230}&\pgfutilensuremath {0.168}&\pgfutilensuremath {0.153}&\pgfutilensuremath {0.211}&\pgfutilensuremath {0.198}&\pgfutilensuremath {0.297}\\%
&BIM&$\bf \pgfutilensuremath {0.002}$&\pgfutilensuremath {0.001}&\pgfutilensuremath {0.005}&\pgfutilensuremath {0.046}&\pgfutilensuremath {0.098}&\pgfutilensuremath {0.066}&\pgfutilensuremath {0.052}&\pgfutilensuremath {0.081}&\pgfutilensuremath {0.086}&\pgfutilensuremath {0.143}\\%
&PGD&\pgfutilensuremath {0.000}&\pgfutilensuremath {0.002}&\pgfutilensuremath {0.003}&\pgfutilensuremath {0.046}&\pgfutilensuremath {0.097}&\pgfutilensuremath {0.061}&\pgfutilensuremath {0.054}&\pgfutilensuremath {0.082}&\pgfutilensuremath {0.090}&\pgfutilensuremath {0.142}\\\bottomrule %
\end {tabular}%

%% file: tables/table.ml.item_user.fair_usravg.tex
\begin {tabular}{rrrrrrrrrrrrrrrrr}%
\toprule &&\multicolumn {5}{c}{Item pop} & \multicolumn {5}{c}{User gender} & \multicolumn {5}{c}{User interactions} \\Model&&$f_0$&$f_1$&$f_4$&MADr&MADR&$f_0$&$f_1$&$f_2$&MADr&MADR&$f_0$&$f_1$&$f_4$&MADr&MADR\\\cmidrule (r){1-2}\cmidrule (rl){3-7}\cmidrule (rl){8-12}\cmidrule (rl){13-17}%
\multirow {4}{*}{\noDef }&initial&$\bf \pgfutilensuremath {-0.483}$&$\bf \pgfutilensuremath {-1.574}$&$\bf \pgfutilensuremath {-0.005}$&\pgfutilensuremath {0.040}&\pgfutilensuremath {0.159}&\pgfutilensuremath {-0.001}&\pgfutilensuremath {-0.109}&\pgfutilensuremath {-0.143}&$\bf \pgfutilensuremath {0.050}$&\pgfutilensuremath {0.003}&\pgfutilensuremath {-0.116}&$\bf \pgfutilensuremath {-0.138}$&\pgfutilensuremath {-1.480}&\pgfutilensuremath {0.618}&\pgfutilensuremath {0.030}\\%
&FGSM&\pgfutilensuremath {-0.929}&\pgfutilensuremath {-3.056}&\pgfutilensuremath {-0.042}&$\bf \pgfutilensuremath {0.029}$&\pgfutilensuremath {0.140}&$\bf \pgfutilensuremath {0.000}$&\pgfutilensuremath {-0.111}&$\bf \pgfutilensuremath {-0.140}$&\pgfutilensuremath {0.067}&\pgfutilensuremath {0.002}&\pgfutilensuremath {-0.110}&\pgfutilensuremath {-0.158}&\pgfutilensuremath {-1.514}&$\bf \pgfutilensuremath {0.614}$&\pgfutilensuremath {0.028}\\%
&BIM&\pgfutilensuremath {-2{,}039.764}&\pgfutilensuremath {-334.326}&\pgfutilensuremath {-326.189}&\pgfutilensuremath {0.066}&$\bf \pgfutilensuremath {0.079}$&\pgfutilensuremath {-0.003}&$\bf \pgfutilensuremath {-0.088}$&\pgfutilensuremath {-0.170}&\pgfutilensuremath {0.373}&\pgfutilensuremath {0.003}&$\bf \pgfutilensuremath {-0.004}$&\pgfutilensuremath {-0.542}&$\bf \pgfutilensuremath {-0.679}$&\pgfutilensuremath {1.781}&$\bf \pgfutilensuremath {0.002}$\\%
&PGD&\pgfutilensuremath {-3{,}167.250}&\pgfutilensuremath {-8{,}615.699}&\pgfutilensuremath {-506.580}&\pgfutilensuremath {0.062}&\pgfutilensuremath {0.083}&\pgfutilensuremath {0.000}&\pgfutilensuremath {-0.111}&\pgfutilensuremath {-0.140}&\pgfutilensuremath {0.234}&$\bf \pgfutilensuremath {0.001}$&\pgfutilensuremath {-0.024}&\pgfutilensuremath {-0.323}&\pgfutilensuremath {-0.910}&\pgfutilensuremath {1.564}&\pgfutilensuremath {0.005}\\%
\cmidrule (r){1-2}\cmidrule (rl){3-7}\cmidrule (rl){8-12}\cmidrule (rl){13-17}\multirow {4}{*}{\Def }&initial&\pgfutilensuremath {-0.147}&$\bf \pgfutilensuremath {-0.576}$&$\bf \pgfutilensuremath {-0.105}$&\pgfutilensuremath {0.225}&\pgfutilensuremath {0.424}&\pgfutilensuremath {-0.001}&\pgfutilensuremath {-0.104}&\pgfutilensuremath {-0.149}&\pgfutilensuremath {0.084}&\pgfutilensuremath {0.004}&\pgfutilensuremath {-0.092}&$\bf \pgfutilensuremath {-0.162}$&\pgfutilensuremath {-1.329}&\pgfutilensuremath {1.995}&\pgfutilensuremath {0.028}\\%
&FGSM&$\bf \pgfutilensuremath {-0.104}$&\pgfutilensuremath {-0.646}&\pgfutilensuremath {-0.121}&\pgfutilensuremath {0.171}&\pgfutilensuremath {0.302}&$\bf \pgfutilensuremath {-0.001}$&\pgfutilensuremath {-0.105}&$\bf \pgfutilensuremath {-0.147}$&$\bf \pgfutilensuremath {0.038}$&$\bf \pgfutilensuremath {0.004}$&\pgfutilensuremath {-0.093}&\pgfutilensuremath {-0.166}&\pgfutilensuremath {-1.403}&$\bf \pgfutilensuremath {1.674}$&\pgfutilensuremath {0.025}\\%
&BIM&\pgfutilensuremath {-3{,}533.378}&\pgfutilensuremath {-9{,}611.568}&\pgfutilensuremath {-565.161}&$\bf \pgfutilensuremath {0.095}$&$\bf \pgfutilensuremath {0.155}$&\pgfutilensuremath {-0.007}&\pgfutilensuremath {-0.074}&\pgfutilensuremath {-0.193}&\pgfutilensuremath {0.302}&\pgfutilensuremath {0.005}&\pgfutilensuremath {-0.014}&\pgfutilensuremath {-0.435}&\pgfutilensuremath {-0.719}&\pgfutilensuremath {4.175}&\pgfutilensuremath {0.005}\\%
&PGD&\pgfutilensuremath {-1{,}543.481}&\pgfutilensuremath {-287.878}&\pgfutilensuremath {-246.845}&\pgfutilensuremath {0.263}&\pgfutilensuremath {0.330}&\pgfutilensuremath {-0.011}&$\bf \pgfutilensuremath {-0.064}$&\pgfutilensuremath {-0.213}&\pgfutilensuremath {0.328}&\pgfutilensuremath {0.006}&$\bf \pgfutilensuremath {-0.010}$&\pgfutilensuremath {-0.426}&$\bf \pgfutilensuremath {-0.702}$&\pgfutilensuremath {4.177}&$\bf \pgfutilensuremath {0.004}$\\\bottomrule %
\end {tabular}%

%% file: tables/table.lastfm.item_user.fair_usravg.tex
\begin {tabular}{rrrrrrrrrrrrrrrrr}%
\toprule &&\multicolumn {5}{c}{Item pop} & \multicolumn {5}{c}{User gender} & \multicolumn {5}{c}{User interactions} \\Model&&$f_0$&$f_1$&$f_4$&MADr&MADR&$f_0$&$f_1$&$f_2$&MADr&MADR&$f_0$&$f_1$&$f_4$&MADr&MADR\\\cmidrule (r){1-2}\cmidrule (rl){3-7}\cmidrule (rl){8-12}\cmidrule (rl){13-17}%
\multirow {4}{*}{\noDef }&initial&\pgfutilensuremath {-1{,}161.806}&\pgfutilensuremath {-4{,}646.677}&\pgfutilensuremath {-185.725}&\pgfutilensuremath {0.120}&\pgfutilensuremath {0.032}&\pgfutilensuremath {-0.016}&\pgfutilensuremath {-0.188}&\pgfutilensuremath {-0.499}&\pgfutilensuremath {0.147}&\pgfutilensuremath {0.051}&\pgfutilensuremath {-0.015}&\pgfutilensuremath {-0.822}&$\bf \pgfutilensuremath {-0.353}$&$\bf \pgfutilensuremath {0.557}$&\pgfutilensuremath {0.051}\\%
&FGSM&\pgfutilensuremath {-397.483}&\pgfutilensuremath {-3{,}094.772}&\pgfutilensuremath {-63.435}&\pgfutilensuremath {0.123}&\pgfutilensuremath {0.033}&\pgfutilensuremath {-0.016}&$\bf \pgfutilensuremath {-0.180}$&\pgfutilensuremath {-0.489}&$\bf \pgfutilensuremath {0.141}$&\pgfutilensuremath {0.031}&$\bf \pgfutilensuremath {-0.008}$&\pgfutilensuremath {-0.730}&\pgfutilensuremath {-0.395}&\pgfutilensuremath {0.686}&\pgfutilensuremath {0.022}\\%
&BIM&\pgfutilensuremath {-46.740}&\pgfutilensuremath {-186.212}&\pgfutilensuremath {-7.314}&\pgfutilensuremath {0.031}&\pgfutilensuremath {0.003}&$\bf \pgfutilensuremath {-0.002}$&\pgfutilensuremath {-0.372}&\pgfutilensuremath {-0.258}&\pgfutilensuremath {0.312}&$\bf \pgfutilensuremath {0.001}$&\pgfutilensuremath {-0.206}&$\bf \pgfutilensuremath {-0.243}$&\pgfutilensuremath {-2.591}&\pgfutilensuremath {3.153}&\pgfutilensuremath {0.005}\\%
&PGD&$\bf \pgfutilensuremath {-20.224}$&$\bf \pgfutilensuremath {-156.603}$&$\bf \pgfutilensuremath {-3.149}$&$\bf \pgfutilensuremath {0.021}$&$\bf \pgfutilensuremath {0.002}$&\pgfutilensuremath {-0.011}&\pgfutilensuremath {-0.480}&$\bf \pgfutilensuremath {-0.243}$&\pgfutilensuremath {0.290}&\pgfutilensuremath {0.001}&\pgfutilensuremath {-0.025}&\pgfutilensuremath {-0.282}&\pgfutilensuremath {-0.694}&\pgfutilensuremath {3.062}&$\bf \pgfutilensuremath {0.002}$\\%
\cmidrule (r){1-2}\cmidrule (rl){3-7}\cmidrule (rl){8-12}\cmidrule (rl){13-17}\multirow {4}{*}{\Def }&initial&\pgfutilensuremath {-747.062}&\pgfutilensuremath {-5{,}853.077}&\pgfutilensuremath {-119.395}&$\bf \pgfutilensuremath {0.468}$&\pgfutilensuremath {0.055}&\pgfutilensuremath {-0.010}&\pgfutilensuremath {-0.190}&\pgfutilensuremath {-0.416}&\pgfutilensuremath {0.224}&\pgfutilensuremath {0.048}&$\bf \pgfutilensuremath {-0.026}$&$\bf \pgfutilensuremath {-0.921}$&\pgfutilensuremath {-0.291}&\pgfutilensuremath {2.057}&\pgfutilensuremath {0.079}\\%
&FGSM&\pgfutilensuremath {-1{,}242.414}&\pgfutilensuremath {-4{,}969.776}&\pgfutilensuremath {-198.632}&\pgfutilensuremath {0.583}&\pgfutilensuremath {0.066}&$\bf \pgfutilensuremath {-0.009}$&\pgfutilensuremath {-0.193}&$\bf \pgfutilensuremath {-0.413}$&$\bf \pgfutilensuremath {0.108}$&\pgfutilensuremath {0.042}&\pgfutilensuremath {-0.028}&\pgfutilensuremath {-0.930}&\pgfutilensuremath {-0.279}&$\bf \pgfutilensuremath {1.060}$&\pgfutilensuremath {0.074}\\%
&BIM&$\bf \pgfutilensuremath {-2.238}$&$\bf \pgfutilensuremath {-8.706}$&$\bf \pgfutilensuremath {-0.217}$&\pgfutilensuremath {0.672}&$\bf \pgfutilensuremath {0.035}$&\pgfutilensuremath {-0.014}&$\bf \pgfutilensuremath {-0.178}$&\pgfutilensuremath {-0.459}&\pgfutilensuremath {0.941}&$\bf \pgfutilensuremath {0.021}$&\pgfutilensuremath {-0.067}&\pgfutilensuremath {-1.257}&$\bf \pgfutilensuremath {-0.200}$&\pgfutilensuremath {6.127}&\pgfutilensuremath {0.046}\\%
&PGD&\pgfutilensuremath {-309.333}&\pgfutilensuremath {-2{,}419.092}&\pgfutilensuremath {-49.342}&\pgfutilensuremath {0.742}&\pgfutilensuremath {0.039}&\pgfutilensuremath {-0.022}&\pgfutilensuremath {-0.200}&\pgfutilensuremath {-0.562}&\pgfutilensuremath {0.978}&\pgfutilensuremath {0.025}&\pgfutilensuremath {-0.063}&\pgfutilensuremath {-1.237}&\pgfutilensuremath {-0.210}&\pgfutilensuremath {7.015}&$\bf \pgfutilensuremath {0.046}$\\\bottomrule %
\end {tabular}%

%% file: 6Conlusion.tex
We proposed iterative adversarial perturbations on the model embedding of matrix factorization-based recommender models. We studied the impact of the proposed perturbations with extensive experiments on two datasets (i.e., \lfm and \ml) and two state-of-the-art MF recommenders, i.e., BPR-MF, and AMF --- an extension of BPR-MF that integrates the adversarial training as the defense against single-step perturbations. 
Our experiments show that under a fixed perturbation budget, the presented multi-step perturbation strategies, namely the basic iterative method (BIM), and projected gradient descent (PGD), are considerably more effective than the state-of-the-art single-step FGSM method. We verified the degradation of recommendation quality along with accuracy and beyond-accuracy metrics. In particular, experiment validations showed that: (i) non-defended recommenders perturbed by the multi-step perturbation strategies can be impaired/weakened so much so that their performance becomes worse than a random recommender and, (ii) even the adversarially defended model against FGSM can lose half of its recommendation performance, --- i.e., after being confronted with an iterative perturbation, they preserve only half of the learned personalized users' preferences. Equivalently, we verified that iterative perturbations could produce the same performance drop as of FGSM perturbations with 5-time smaller perturbation levels. These results evidence the vulnerability of the personalized BPR-learned models, both in defended and non-defended scenarios.

Then, we investigated the impact of structural dataset characteristics (i.e., density, size, and shape) on the efficacy of adversarial perturbations and verified that recommenders trained on denser datasets are more robust to adversarial perturbations while increasing the shape, or the size, the model becomes more vulnerable.
The challenges that we plan to investigate in the future are on the study of defense strategies to robustify the recommender against the iterative perturbations in order to make RS robust to slight variations of model parameters. 
Moreover, we intend to investigate the fairness of the recommendations, by exploring users and items clusters (for instance, created based on their attributes or according to their popularity) and investigate methods to find the impact on fairness metrics based on the perturbations of the recommender parameters.